\documentclass[preprint]{aastex}

\usepackage{lscape,natbib,amssymb}
\newcommand{\amm}{NH$_3$}
\newcommand{\dia}{N$_2$H$^+$}

\newcommand{\kms}{\,km\,s$^{-1}$}
\newcommand{\cc}{\,cm$^{-3}$}
\newcommand{\one}{Paper I}
\newcommand{\edits}{}

\shorttitle{\dia\, in Ophiuchus}
\shortauthors{Friesen et al.}

\begin{document}

\title{The Initial Conditions of Clustered Star Formation II: \\ \dia\, Observations of the Ophiuchus B Core}
\author{R. K. Friesen\altaffilmark{1,2}}
\altaffiltext{1}{Department of Physics and Astronomy, University of Victoria, PO Box 3055, STN CSC, Victoria BC CANADA V8W 3P6}
\email{rachel.friesen@nrc-cnrc.gc.ca}

\author{J. Di Francesco\altaffilmark{2,1}}
\altaffiltext{2}{National Research Council Canada, Herzberg Institute of Astrophysics, 5071 West Saanich Road, Victoria, British Columbia, Canada V9E 2E7}

\author{Y. Shimajiri\altaffilmark{3}}
\altaffiltext{3}{Department of Astronomy, School of Science, University of Tokyo, Bunkyo, Tokyo 113-0033, Japan}

\and

\author{S. Takakuwa\altaffilmark{4}}
\altaffiltext{4}{Academia Sinica Institute of Astronomy and Astrophysics, P.O. Box 23-141, Taipei 106, Taiwan}

\begin{abstract}
We present a Nobeyama 45\,m Radio Telescope map and Australia Telescope Compact Array pointed observations of \dia\, 1-0 emission towards the clustered, low mass star forming Oph B Core within the Ophiuchus molecular cloud. We compare these data with previously published results of high resolution \amm\, (1,1) and (2,2) observations in Oph B. We use 3D {\sc clumpfind} to identify emission features in the single-dish \dia\, map, and find that the \dia\, `clumps' match well similar features previously identified in \amm\, (1,1) emission, but are frequently offset to clumps identified at similar resolution in 850\,\micron\, continuum emission. Wide line widths in the Oph B2 sub-Core indicate non-thermal motions dominate the Core kinematics, and remain transonic at densities $n \sim 3 \times 10^5$\,\cc\, with large scatter and no trend with $N(\mbox{H$_2$})$. In contrast, non-thermal motions in Oph B1 and B3 are subsonic with little variation, but also show no trend with H$_2$ column density. Over all Oph B, non-thermal \dia\, line widths are substantially narrower than those traced by \amm, making it unlikely \amm\, and \dia\, trace the same material, but the $v_{\mbox{\tiny{LSR}}}$ of both species agree well. We find evidence for accretion in Oph B1 from the surrounding ambient gas. The \amm\,/\,\dia\, abundance ratio is larger towards starless Oph B1 than towards protostellar Oph B2, similar to recent observational results in other star-forming regions. The interferometer observations reveal small-scale structure in \dia\, 1-0 emission, which are again offset from continuum emission. No interferometric \dia\, emission peaks were found to be coincident with continuum clumps. In particular, the $\sim 1$\,M$_\odot$ B2-MM8 clump is associated with a \dia\, emission minimum and surrounded by a broken ring-like \dia\, emission structure, suggestive of \dia\, depletion. We find a strong general trend of decreasing \dia\, abundance with increasing $N(\mbox{H$_2$})$ in Oph B which matches that found for \amm. 
\end{abstract}


\section{Introduction}

Stars form out of the gravitational collapse of centrally condensed cores of dense molecular gas. Recent years have seen leaps forward in our understanding of the structure and evolution of isolated, star forming cores. Most star formation, however, occurs in clustered environments \citep{lada03}. These regions are more complex, with complicated observed geometries, and contain cores which tend to have higher densities and more compact sizes than those found in isolation \citep{wardthompson07}. It is likely that due to these differences the evolution of filaments and cores in clustered regions proceeds differently than in the isolated cases. Characterizing the physical and chemical structures of these more complicated regions are thus the first steps towards a better understanding of the process of clustered star formation. 

It is now clear that molecular cores become extremely chemically differentiated, as many molecules commonly used for tracing molecular gas, such as CO, become severely depleted in the innermost core regions through adsorption onto dust grains \citep[see, e.g.,][for a review]{difran07}.  In a recent paper \citep[hereafter \one]{friesen09}, we studied the dense gas in several cluster forming Cores\footnotemark\footnotetext{\edits{In this paper, we describe Oph A, B, C, etc., as `Cores' since this is how these features were named in DCO$^+$ observations of the L1688 region by \citet{loren90}. Since then, higher-resolution data such as those described in this paper have revealed substructure in these features that could be themselves precursors to stars, i.e., cores.  To avoid confusion, we refer the larger features in Oph as `Cores' and Core substructure identified by {\sc clumpfind} as `clumps'.}} in the Ophiuchus molecular cloud (Oph B, C and F) through high resolution observations of \amm\, (1,1) and (2,2) emission made at the Green Bank Telescope (GBT), the Australia Telescope Compact Array (ATCA) and the Very Large Array (VLA). The Ophiuchus molecular cloud \citep[$\sim 120$\,pc distant; ][]{loinard08,lombardi08,knude98} is our closest example of ongoing, clustered star formation. We found that the Ophiuchus Cores presented physical characteristics similar to those found in studies of both isolated and clustered environments. In Oph C, for example, gas motions are subsonic (mean $\sigma_v = 0.16$\,\kms), and decrease in magnitude along with the kinetic gas temperature ($T_K$) towards the thermal dust continuum emission peak in a manner reminiscent of findings in the isolated cores L1544 \citep{crapsi07}; etc. In contrast to Oph C, \amm\, line widths in Oph B are dominated by transonic non-thermal motions and the gas temperatures are warmer than typically found in isolated regions ($\langle T_K \rangle = 15$\,K) and nearly constant across the Core. No contrast in any parameters [such as $T_K$, ratio of non-thermal to thermal line width $\sigma_{\mbox{\tiny{NT}}}\,/\,c_s$, or \amm\, column density $N(\mbox{\amm})$] was found between the two main components of Oph B, labelled B1 and B2, despite the presence of several embedded Class I protostars in B2 while B1 appears starless. Additionally, on small scales ($\sim 15$\arcsec, or 1800\,AU at 120\,pc), significant offsets were found between the peaks of integrated \amm\, (1,1) intensity and those of submillimeter continuum emission, as well as between individual `clumps' identified through 3D {\sc clumpfind} \citep{williams94} in the \amm\, data cube and those found through 2D {\sc clumpfind} in 850\,\micron\, continuum emission. Finally, evidence was found for a decreasing fractional \amm\, abundance with increasing H$_2$ column density (as traced by 850\,\micron\, continuum emission), suggesting that \amm\, (1,1) may not be tracing the densest gas in the Oph Cores. 

The diazenylium ion, \dia, has been shown to be a preferential tracer of quiescent dense gas in molecular clouds \citep{womack92,caselli02,tafalla02,tafalla04,difran04}. The 1-0 rotational transition has a critical density $n_{cr} \sim 2 \times 10^5$\,\cc, $\sim 10-100 \times$ greater than that of \amm\, (1,1), and is therefore possibly better suited to probing gas properties at the high densities expected in clustered star forming Cores. Based on both observations and chemical models of starless cores, \dia\, also appears resilient to depletion at the high densities ($n \gtrsim 10^5$\,\cc) and cold temperatures ($T \sim 10$\,K) characteristic of the later stages of prestellar core evolution \citep[see, e.g.,][]{tafalla04,bergin02}, and is thus expected to be an excellent tracer of the physical conditions of dense cores and clumps. 

Recently, \citet{andre07} observed the Oph A, B, C, E and F Cores in \dia\, 1-0 emission at 26\arcsec\, angular resolution using the IRAM 30\,m telescope. They found that the gas traced by \dia\, emission near clumps identified in millimeter continuum emission \citep{motte98} is characterized, on average, with relatively narrow line widths showing small non-thermal velocity dispersions, in contrast to the large non-thermal motions traced by \amm\, emission in \one. The authors derived virial masses from the \dia\, emission, and found they generally agreed within a factor $\sim 2$ with the masses derived from dust emission, suggesting that the clumps are gravitationally bound and prestellar. The relative motions of the clumps are also small and subvirial, with a crossing time larger than the expected lifetime of the objects, such that clump-clump interactions are not expected to impact their evolution. These results suggest that a dynamic picture of clump evolution involving competitive accretion at the prestellar stage does not accurately describe the star formation process in central Ophiuchus. 

In this work, we discuss the results of higher resolution (18\arcsec, or $\sim 2200$\,AU at 120\,pc) observations of \dia\, 1-0 in the Ophiuchus B Core that reveal the distribution, kinematics and abundance pattern of the Core and associated embedded clumps on smaller physical scales. We additionally examine \dia\, 1-0 structure at even higher resolution ($8\arcsec \times 5\arcsec$), through observations made with the ATCA, towards five locations in Oph B2 where small-scale structure was found in \amm\, (\one) and in previously unpublished Berkeley-Illinois-Maryland Association (BIMA) \dia\, 1-0 observations. We compare these data with our recently published analysis of \amm\, emission in Oph B, in particular focusing on the kinematics and relative abundances of the two species to compare with physical and chemical models of dense core formation and collapse. 

We discuss the observations and calibration and \S2. In \S3, we present the data, and discuss the results of the hyperfine line structure fitting procedure and derivations of the column density, $N(\mbox{\dia})$, and fractional \dia\, abundance, $X(\mbox{\dia})$, in \S4. In \S5, we discuss general trends in the data, and compare these results with those found in \one\, and with studies of dense cores in isolated environments. We summarize our findings in \S6. 



\section{Observations and Data Reduction}

\subsection{Nobeyama 45\,m Radio Telescope}

Single-dish mapping observations of Oph B in \dia\, 1-0 (rest frequency 93.174\,GHz) were performed May 10 - 15, 2007 using the 25-BEam Array Receiver System \citep[BEARS, ][]{sunada00,yamaguchi00} at the Nobeyama 45\,m Radio Telescope. BEARS is a $5 \times 5$ SIS heterodyne receiver array. The autocorrelator spectrometer \citep{sorai00} with 8\,MHz bandwidth and 1024 channels was used as the backend, giving a spectral resolution of 7.8\,kHz or 0.025\kms\, at 93\,GHz. The final spectral resolution was binned to 0.05\kms.  The half-power beamwidth was $\sim 18$\arcsec\, and the main-beam efficiency at 93\,GHz is estimated to be $\eta_{MB} = 0.505$, interpolated from observatory measurements at 86\,GHz and 100\,GHz, with an uncertainty of $\sim 8-10$\,\%.  The beams are separated by 41.1\arcsec\, on the sky. 

Observations were performed in position-switching OTF mode, alternating scanning in Right Ascension (R.A.) and Declination (decl.) to ensure minimal striping artifacts. The array was rotated by an angle of 7\arcdeg\, such that the spacing of individual detectors was 5\arcsec\, in R.A. or decl., and scans of the array were spaced by 25\arcsec. A ten minute integration on the off-source position (+5\arcdeg\, in R.A.) was performed to ensure no emission was present. The weather was only moderately variable over the observations, and most data were taken with $T_{sys} \sim 300 - 450$\,K (DSB). Telescope pointing was checked every 90 minutes on nearby SiO maser sources AF Sco and RS Lib. Typical pointing offsets were $1 - 3$\arcsec. 

Calibration of BEARS was done by observing a bright \dia\, target first with the S100 single sideband receiver, and second with each individual BEARS beam. These data provided a flux scaling factor for each beam to correct for the telescope gain and DSB ratio. The starless core L183 was observed with S100 and the central BEARS beam, but the line emission was not strong enough to use as a calibrator for each BEARS beam. The Oph A core SM1 was then observed with each BEARS beam and the scaling factors for each beam were determined using both sources. The average scaling factor was 1.35, with a range from 1.07 to 1.66.

The scaling factors were applied to the data and a linear baseline was fit to and subtracted from the spectra. The R.A. and decl. scans were gridded separately and then combined using a basket-weave method to minimize baseline drift effects \citep{emerson88}. 

\subsection{BIMA}

Observations of \dia\, 1-0 across the Oph B2 Core were made with the BIMA array located at Hat Creek, CA, U.S.A.  Four tracks were observed on 2001 March 18, 21, and 26 and 2001 May 23, when the array was in its C-configuration.  Observations of Oph B2 consisted of a serial sequence of 12 overlapping positions with centers spaced by 1.06\arcmin, to obtain a Nyquist-sampled mosaic across Oph B2 at 93.2\,GHz.  The correlator was configured to provide 12.5 MHz of bandwidth in one correlator window at 93.2\,GHz, providing 512 channels of 0.04\kms\, width.  The radio source 3C279 was observed at the start of each track to provide passband and flux calibration.  Neptune was observed at the end of each track to provide flux calibration.  The radio source 1733-130 was observed between every set of Oph B2 observations to provide phase calibration.  Values of $T_{sys}$ during the tracks were typically 250-500\,K for all antennas, except antenna 2 where $T_{sys}$ values were a factor of $\sim 1.5$ higher.

The Oph B2 BIMA data were reduced using standard tasks in the MIRIAD package \citep{sault95}. To improve sensitivity to diffuse emission and produce a rounder beam, the visibilities were tapered with a Gaussian 10\arcsec\, FWHM in R.A. and 0.1\arcsec\, FWHM in decl. For the \dia\, data, the final beam size achieved was 19.3\arcsec\, $\times$ 10.6\arcsec\, FHWM (P.A. -12.4\arcdeg) and the resulting sensitivity (1 $\sigma$ rms) per channel was 0.72\,Jy\,beam$^{-1}$, or equivalently 0.49 K.  Only faint, diffuse \dia\, emission was detected towards locations in eastern Oph B2, with even weaker emission spots towards central Oph B2. The faint \dia\, emission was not detected significantly enough to fit the hyperfine line components, but the locations of significant integrated intensity provided targets for later, more sensitive data of Oph B2 with the ATCA (see below).  


\subsection{ATCA}

A total of 11 pointings were observed in \dia\, 1-0 emission at the ATCA towards five targets in July 2006 in the interferometer's most compact (H75) configuration. The BIMA data were used as a guide for the pointings. Figure \ref{fig:nobeyama}a shows the positions of all 11 pointings over an 850\,\micron\, continuum map of Oph B. Table \ref{tab:atca} lists the positions of the targets, the rms noise achieved in Jy\,beam$^{-1}$, and any associated \amm\, clumps identified in \one, continuum objects, or embedded Young Stellar Objects (YSOs). Two of the targets were imaged using mosaics of three Nyquist-spaced pointings each. Three pointings towards the Oph B2 continuum emission peak overlapped, but were not Nyquist-spaced, while the remaining two targets were observed with a single pointing each. An 8\,MHz band width with 512 channels was used for the observations, providing a spectral resolution of 15.6\,kHz or 0.05\,\kms\, at 93.2\,GHz while allowing all hyperfine components of the transition to fit within the band. Baselines ranged from 31\,m to 89\,m for five antennas (the sixth antenna is located 6\,km from the main group and was not used for these observations due to poor phase stability), corresponding to 10.3\,k$\lambda$ - 29.7\,k$\lambda$. The primary beam (field-of-view) was 34\arcsec\, in diameter. 

Observations were made in frequency-switching mode. Observations cycled through each individual pointing between phase calibrator observations to maximize $u,v$ coverage and minimize phase errors. The phase calibrator, 1622-297, was observed every 20 minutes, and was also used to check pointing every hour. Flux and bandpass calibration observations were performed each shift on Mars or Uranus. 

The ATCA data were reduced using the Miriad data reduction package \citep{sault95}. The data were first flagged to remove target observations unbracketed by phase calibrator measurements, data with poor phase stability or anomalous amplitude measurements. The bandpass, gains and phase calibrations were applied, and the data were then deconvolved (or jointly deconvolved, in the case of the mosaiced pointings).  First, the data were transformed from the $u,v$ plane into the image plane using the task INVERT and natural weighting to maximize sensitivity. The data were then deconvolved and restored using the tasks MOSSDI and RESTOR to remove the beam response from the image. MOSSDI uses a Steer-Dewdney-Ito CLEAN algorithm \citep{steer84}. The cleaning limit was set at twice the rms noise level in the beam overlap region for each object. Clean boxes were used to avoid cleaning noise in the outer regions with less beam overlap. The final synthesized beam had a FWHM of $8\arcsec\, \times 5$\arcsec. 

We next convolved the ATCA data only to a final beam FWHM of 18\arcsec\, to match the Nobeyama data and estimate the amount of flux recovered by the interferometer. With the exception of the single pointing towards the \amm\, clump B2-A7, we found the maximum peak intensities in the convolved maps are only $\sim 15 - 20$\,\% of that in the single-dish map, suggesting that the interferometer data alone are only sensitive to $\sim 15 - 20$\,\% of the total \dia\, emission. At the B2-A7 location, the single detected interferometer emission peak is $\gtrsim 50$\,\% of that seen by the single-dish telescope.

\section{Results}

\subsection{Single-Dish Data}

\begin{figure}
\plotone{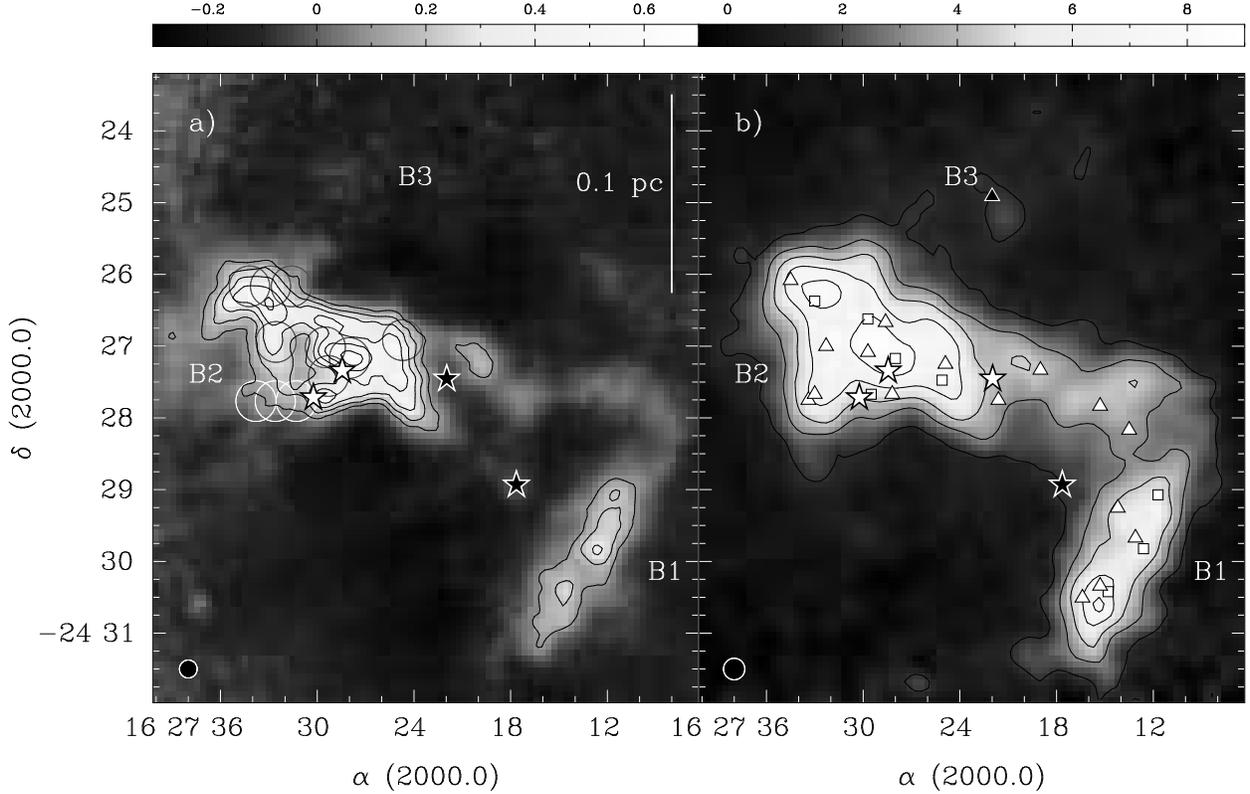}
\caption{a) 850\,\micron\, continuum emission in the Ophiuchus B2 core in Jy\,beam$^{-1}$, first mapped by \citet{johnstone00} at the JCMT. Black contours begin at 0.1\,Jy\,beam$^{-1}$ and increase by 0.1\,Jy\,beam$^{-1}$. The 15\arcsec\, beam is shown at lower left. Grey and white circles show locations of ATCA observations with a primary beam diameter of 34\arcsec. In both plots, stars indicate the positions of Class I objects \citep{enoch08}.  b) The Ophiuchus B core in \dia\, 1-0 integrated intensity in K\,\kms\, ($T_A^*$) observed at Nobeyama. Oph B1, B2 and B3 are labelled. (Oph B3 is not prominent in integrated intensity due to the extremely narrow lines found at this location) Contours begin at 1.5\,K\,\kms\, and increase by 1.5\,K\,\kms. The 18\arcsec\, beam is shown at lower left. Triangles indicate peak positions of \dia\, 1-0 clumps discussed in \S\ref{sec:clumps} and listed in Table \ref{tab:clump} while square indicate locations of 850\,\micron\, continuum clumps \citep{jorgensen08}.}
\label{fig:nobeyama}
\end{figure}

In Figure \ref{fig:nobeyama}a, we show 850\,\micron\, continuum emission in Oph B at 15\arcsec\, FWHM resolution, first mapped with the Submillimetre Common User Bolometer Array (SCUBA) at the James Clerk Maxwell Telescope (JCMT) by \citet{johnstone00} and recently re-reduced and combined with all other SCUBA archive data in the region by \citet{jorgensen08}, following the description in \citet{kirk06}. The Oph B Core can be separated into three distinct sub-Cores, B1, B2, and B3, which are labelled on the Figure. We also show the locations of four Class I protostars detected and classified through Spitzer infrared observations \citep{enoch08}, some of which were previously identified as YSOs \citep{vssg,elias78}. 

Figure \ref{fig:nobeyama}b shows the \dia\, 1-0 intensity in the Oph B Core at 18\arcsec\, FWHM resolution, integrated over only those velocity channels containing emission from the hyperfine components. On 18\arcsec\, scales, the integrated \dia\, intensity within Oph B1 and B2 is remarkably smooth and shows several maxima on scales significantly larger than the beam. On average, the integrated \dia\, intensity closely follows the continuum emission in the Core, with a few exceptions. First, we find significant \dia\, emission in southeast Oph B2, where little continuum emission is seen. In fact, the entire eastern half of B2 contains strong \dia\, emission, while the continuum emission becomes significantly fainter to the east of  the central B2 850\,\micron\, emission peak. Second, the integrated \dia\, intensity maximum in B2 is offset from the continuum maximum by $\sim 23$\arcsec, although strong \dia\, emission is also found at the continuum peak. Third, the continuum and \dia\, integrated intensity peaks in Oph B1 are also offset. Fourth, the filament connecting Oph B1 and B2 is more prominent in \dia\, than in continuum emission. The Oph B3 Core is undetected in 850\,\micron\, continuum. Although it is a clear feature in the \dia\, data, it is faint in \dia\, integrated intensity due to the extremely narrow line widths associated with the Core (discussed further in \S\ref{sec:analysis}). 

\subsubsection{Identifying discrete \dia\, `clumps'}
\label{sec:clumps}

We next identify discrete emission `clumps' in the Nobeyama \dia\, 1-0 emission data cube using the 3D version of the automated structure-finding routine {\sc clumpfind} \citep{williams94}. {\sc clumpfind} uses specified brightness contour intervals to search through the data cube for distinct \edits{emission features} identified by closed contours in position and velocity. The size of the structures found are determined by including adjacent pixels down to an intensity threshold, or until the outer edges of two separate structures meet. The clump location is defined as the emission peak within the core. \edits{The identification of emission clumps in the data allows us to discuss the absolute locations of \dia\, emission within Oph B with respect to the locations of objects identified in continuum emission, as well as emission features in other molecular line tracers. We caution that the identification of \dia\, `clumps' in Oph B does not imply any association with underlying physical structure in the core.} 

The rms noise level in the data cube is essentially constant (rms $\sim 0.08$\,K in $T_A^*$ units) over most of the region observed, but emission in Oph B1 extends near the edge of the area of best OTF mapping overlap and consequently has slightly lower sensitivity. We therefore created a signal-to-noise data cube by dividing the spectrum at each pixel by the rms noise in the off-line spectral channels. We then ran {\sc clumpfind} specifying brightness contours of $4 \times$ the S/N ratio with an intensity threshold of $3 \times$ the data rms. The strongest hyperfine components of the \dia\, 1-0 line overlap significantly over much of the region observed, so we ran the algorithm on the isolated $F_1F \rightarrow F'_1F' = 0 1 \rightarrow 1 2$ component only. This strategy allowed the clear separation of multiple velocity components along the line of sight, but reduced the signal-to-noise ratio of the detections since the isolated component can be substantially fainter (up to a factor of $\sim 2.3$) than the central hyperfine emission. We therefore performed several checks to ensure identified \dia\, objects were legitimate. We removed any identified clumps with areas less than the Nobeyama beam, clumps with a peak location at the edge of the mapped region, and clumps with reported line widths less than twice the velocity resolution of the observations (0.05\,\kms). These limits culled most of the spurious objects, and we finally removed several additional objects through visual inspection. 

We found a total of 17 separate \dia\, clumps in Oph B from the Nobeyama data. Table \ref{tab:clump} lists the locations and FWHM in AU of the clumps, as well as the peak temperatures ($T_A^*$) of the isolated and main hyperfine \dia\, components, and the signal-to-noise ratio of the isolated hyperfine component. We plot the positions of the clumps, i.e., the positions of peak brightness temperature of the isolated component, in Figure \ref{fig:nobeyama}b. The centroid velocities and line widths of the \dia\, clumps are also returned by {\sc clumpfind}, but we determined these values instead through spectral line fitting (described in further detail in \S3). The line fitting algorithm provided more accurate results given the hyperfine structure of the \dia\, 1-0 line. 

\begin{figure}
\epsscale{0.8}
\plotone{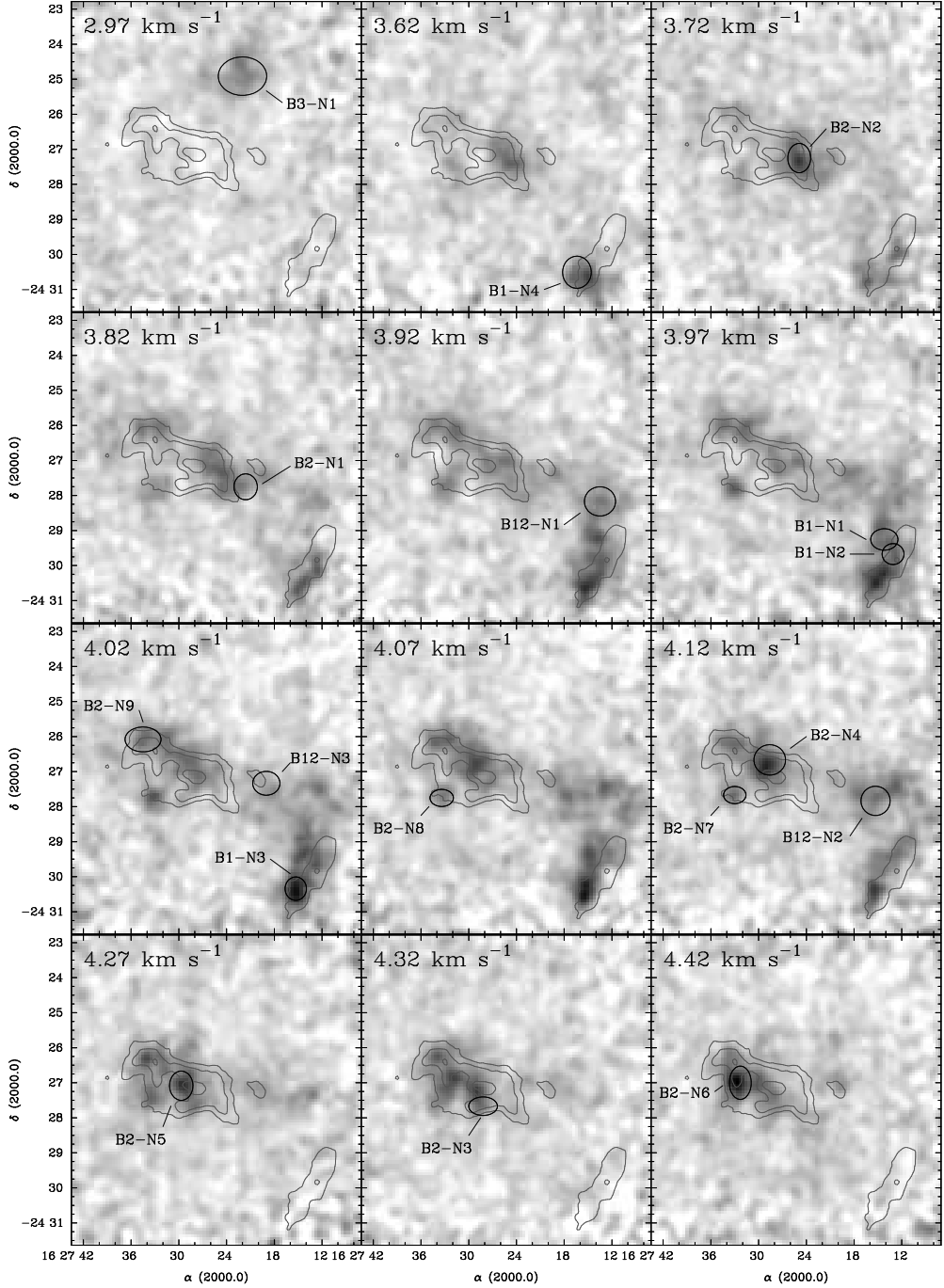}
\caption{\dia\, 1-0 channel maps showing the locations and {\sc clumpfind}-determined FWHM in R.A. and Decl. of the \dia\, clumps at the velocity channel closest to their fitted $v_{\mbox{\tiny{LSR}}}$ (where the velocity channel spacing is 0.05\,\kms). Greyscale ranges from -0.3\,K to 1.9\,K ($T_{MB}$. In locations where multiple velocity components were found, clumps are shown at the fitted $v_{\mbox{\tiny{LSR}}}$ most closely matching the {\sc clumpfind}-determined velocity at the peak line intensity. Grey contours show 850\,\micron\, continuum emission at 0.1, 0.3 and 0.5\,Jy\,beam$^{-1}$ (15\arcsec\, FWHM).}
\label{fig:channel_maps}
\epsscale{1}
\end{figure}

\edits{We show in Figure \ref{fig:channel_maps} the \dia\, clumps as ellipses overlaid on the emission at the velocity channels that most closely match the clumps' centroid velocities determined through spectral line fitting in \S3. The ellipse axes are the returned clump FWHMs in R.A. and decl. The 850\,\micron\, continuum emission contours are also shown in Figure \ref{fig:channel_maps}. In many cases, an offset between the \dia\, clumps and continuum emission is clear. }

\begin{figure}
\plotone{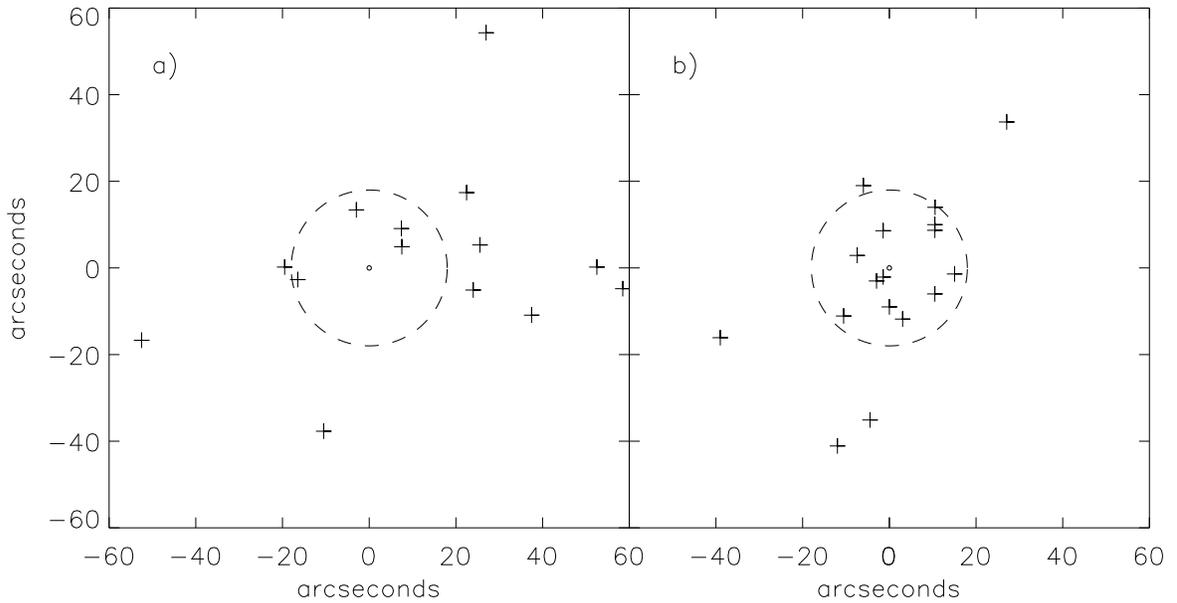}
\caption{a) Minimum angular separation in R.A. and decl. between individual \dia\, 1-0 clumps identified through {\sc clumpfind} and 850\,\micron\, continuum clumps identified by \citet{jorgensen08}. In both plots, the dashed circle has a radius matching the 18\arcsec\, FWHM of the \dia\, 1-0 observations.  On average, \dia\, clumps in Oph B show a positional offset from continuum clumps similar to that found for clumps identified in \amm\, (1,1) emission in \one. b) Minimum angular separation in R.A. and decl. between individual \dia\, 1-0 clumps and \amm\, (1,1) clumps identified in \one. Good correlation is found between \dia\, and \amm\, objects in Oph B.}
\label{fig:b_sep}
\end{figure}

We first compared the locations of the \dia\, clumps with the peak locations of 850\,\micron\, continuum clumps identified by \citet{jorgensen08} in the SCUBA data. Figure \ref{fig:b_sep}a shows the angular separation between \dia\, clumps and the peak location of the nearest 850\,\micron\, continuum clump. Few continuum clumps lie within a radius of 18\arcsec\, in angular distance from an \dia\, clump, suggesting that many of the \dia\, clumps thus identified are not correlated with continuum objects. This result is similar to our findings in \one, where a similar analysis of \amm\, emission in Oph B revealed significant offsets between locations of \amm\, clumps and continuum clumps. Interestingly, the mean minimum separation between \dia\, and continuum clumps in Oph B1 ($\sim 20$\arcsec) is half that in Oph B2 ($\sim 40$\arcsec), suggesting there is a better correlation between \dia\, and continuum emission in Oph B1. Furthermore, Figure \ref{fig:b_sep}b shows the angular separation between \dia\, clumps and nearest \amm\, clumps from \one, and shows a clear correlation between the two sets of objects. In the majority of cases, \dia\, and \amm\, clumps are coincident within the resolution of the \dia\, data. 

\edits{Other searches for submillimeter continuum emission clumps in Oph B using different clump-finding algorithms produce different numbers of objects within each sub-Core. For example, the clump catalogue by \citet{simpson08} was created from a combination of all 850\,\micron\, SCUBA data in Ophiuchus, following the technique described by \citet{nutter07}, and contains a larger number of objects within the Oph B Core. Comparison of the \dia\, clumps with the \citeauthor{simpson08} continuum clumps shows a better correspondence than with the \citeauthor{jorgensen08} catalogue, but the median offset between \dia\, and continuum clump locations is still greater than the 18\arcsec\, FWHM beam of the \dia\, observations. In the following, we restrict our analysis to the \citeauthor{jorgensen08} catalogue as we use the same clump identification technique, but note that offsets are also seen between the \dia\, clumps presented here and other continuum clump catalogues.}

We further compared the locations of the \dia\, clumps thus determined with \dia\, clumps identified by \citet{andre07} in \dia\, 1-0 emission observed with the IRAM 30\,m telescope ($\sim 26$\arcsec\, half power beam width). We find good agreement with \citeauthor{andre07} for most of the objects despite their use of a spatial filtering scheme rather than the {\sc clumpfind} algorithm to identify clumps. 

\subsection{ATCA Data}
\label{sec:results-atca}

\begin{figure}
\epsscale{0.45}
\plotone{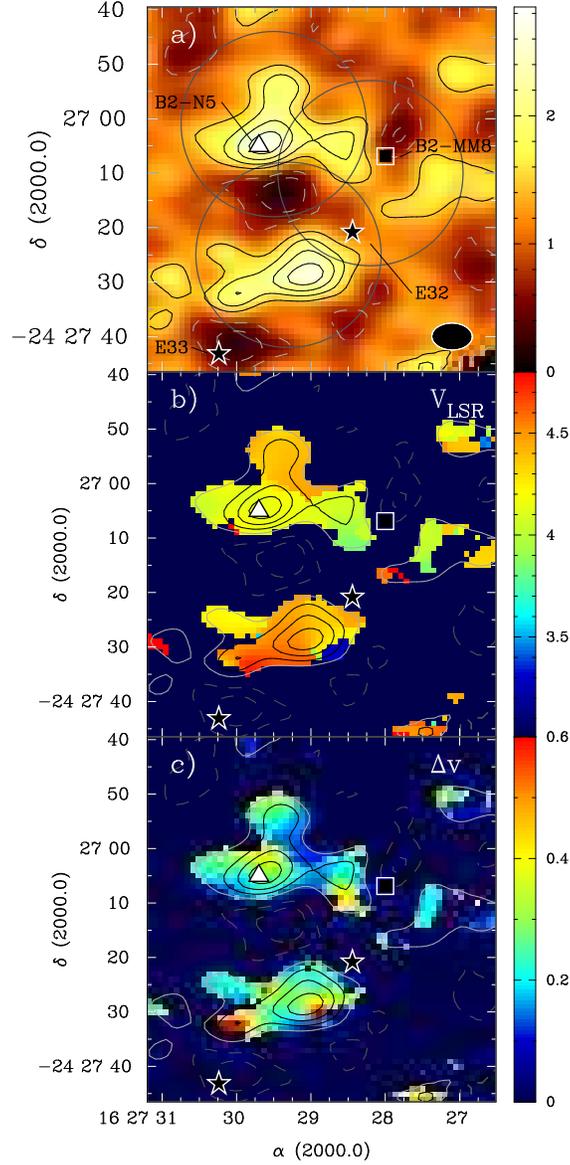}
\caption{a) \dia\, 1-0 integrated intensity in K\,\kms\, ($T_B$) observed at the ATCA. White circles show the primary field of view (34\arcsec) of the ATCA at 93\,GHz for each of the three pointings included in this mosaic. The $\sim 8$\arcsec\, $\times$ 5\arcsec\, FWHM synthesized beam is shown at lower right. In all plots, black contours begin at 0.6\,K\,\kms\, and increase by 0.6\,K\,\kms. Light grey contours show negative integrated intensity at $-0.6$\,K\,\kms, $-1.2$\,K\,\kms\, and $-1.8$\,K\,\kms. The \dia\, clump B2-N5, the continuum clump B2-MM8 \citep{motte98}, and two Class I protostars Elias 32 and 33 \citep{elias78} are labelled. b) High resolution $v_{\mbox{\tiny{LSR}}}$ near B2-MM8. Colour scale is in \kms. c) Fitted $\Delta v$ near MM8. Colour scale is in \kms. }
\label{fig:mm8}
\end{figure}

\begin{figure}
\epsscale{0.5}
\plotone{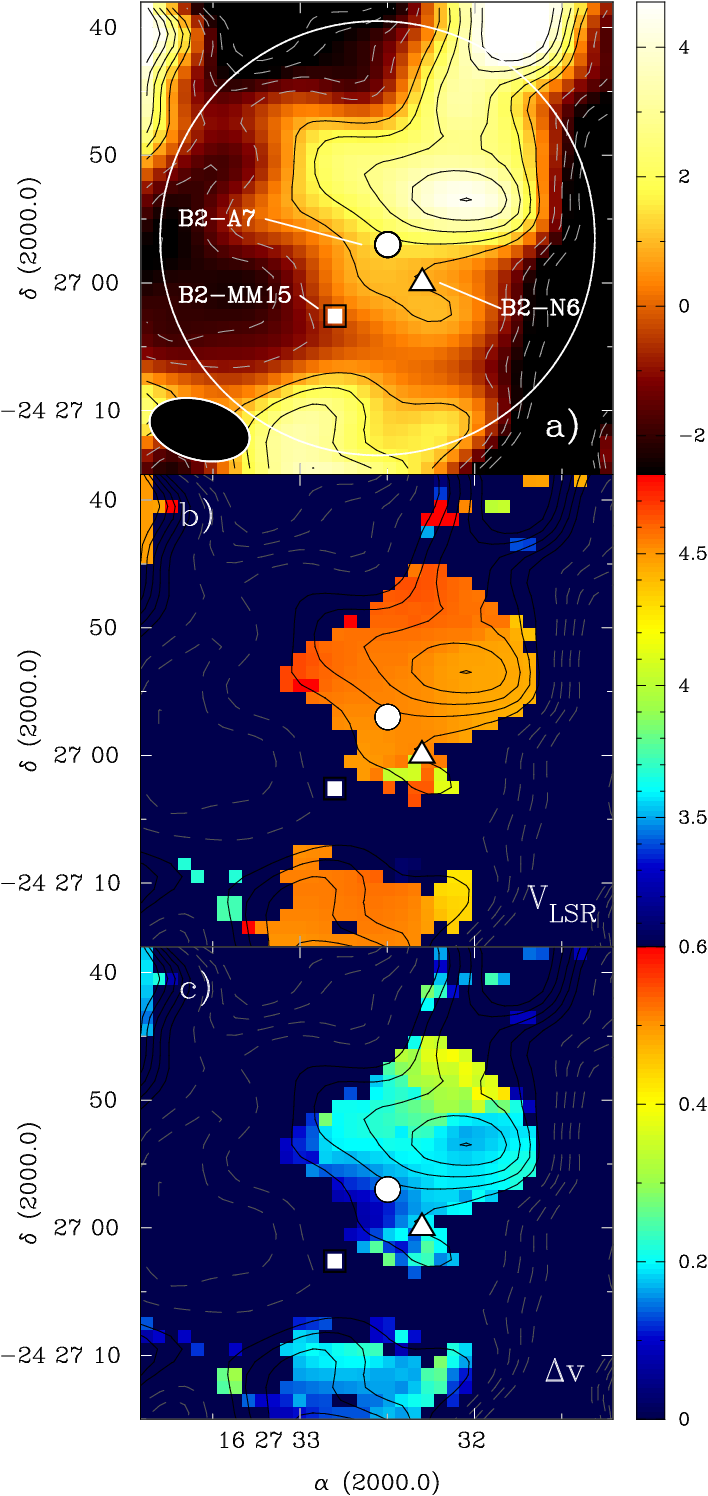}
\caption{a) Single pointing \dia\, 1-0 integrated intensity in the east Ophiuchus B2 core towards the \amm\, (1,1) object B2-A7 identified by \citet{friesen09}.  Colour scale in K\,\kms\, ($T_B$) observed at the ATCA. In all plots, black contours begin at 0.9\,K\,\kms\, and increase by 0.9\,K\,\kms. Light grey contours show negative integrated intensity at $-0.9$\,K\,\kms\, and $-1.8$\,K\,\kms. The 34\arcsec\, diameter white circle shows the primary beam of the ATCA at 93\,GHz. The $\sim 8$\arcsec\, $\times$ 5\arcsec\, FWHM synthesized beam is shown at lower right. The \dia\, clump B2-N6, \amm\, clump B2-A7 (\one) and continuum clump B2-MM15 \citep{motte98} are labelled. b) Line velocity or $v_{\mbox{\tiny{LSR}}}$. Colour scale is in \kms. c) Fitted $\Delta v$ at B2-N6. Colour scale is in \kms. }
\label{fig:nh3peak}
\end{figure}

\begin{figure}
\epsscale{0.6}
\plotone{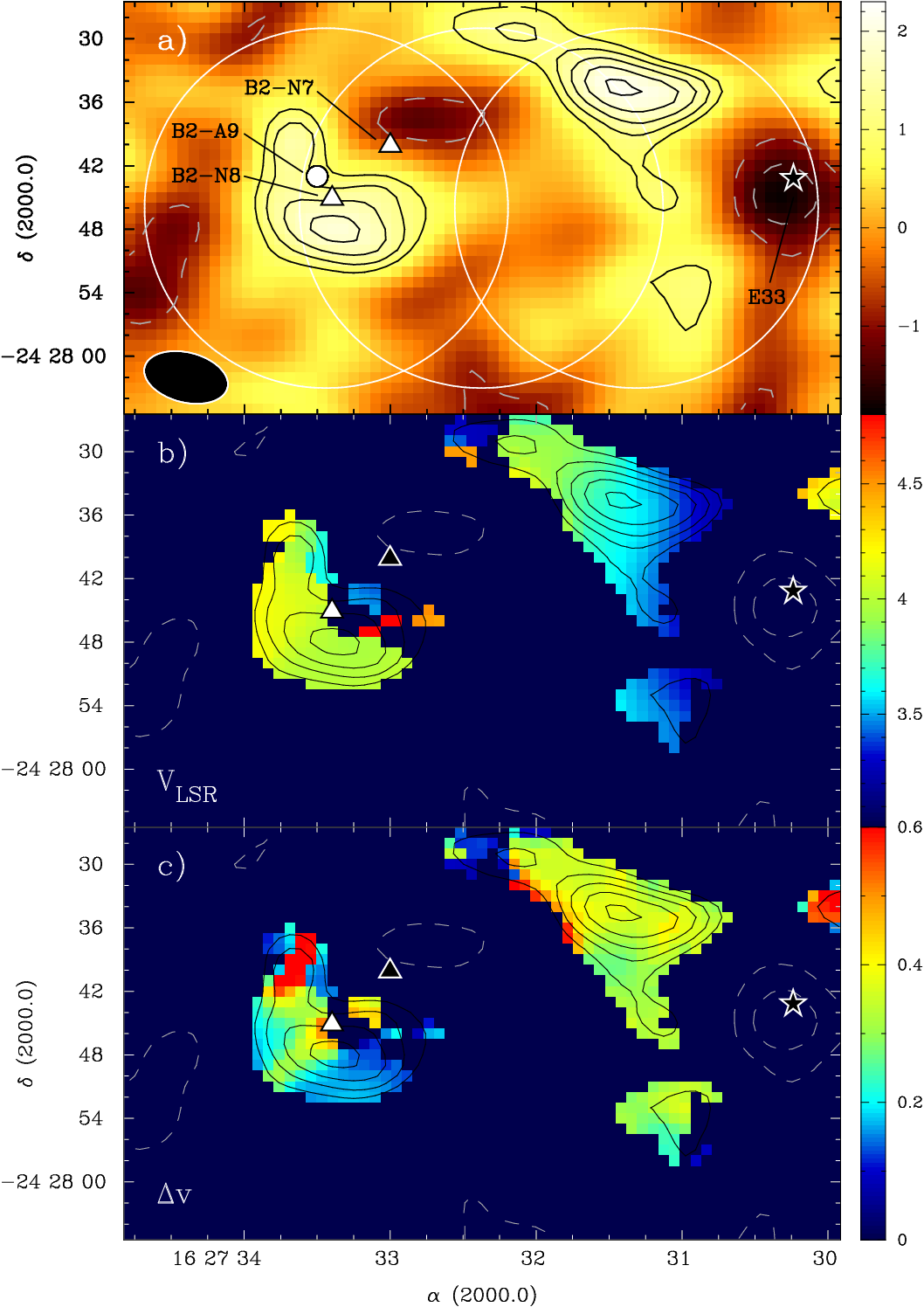}
\caption{a) \dia\, 1-0 integrated intensity in K\,\kms\, ($T_B$) observed at the ATCA in the southeast edge of the Ophiuchus B2 core. White circles show the primary field of view (34\arcsec) of the ATCA at 93\,GHz for each of the three pointings included in this mosaic. The $\sim 8$\arcsec\, $\times$ 5\arcsec\, FWHM synthesized beam is shown at lower right.  In all plots, black contours begin at 0.9\,K\,\kms\, and increase by 0.3\,K\,\kms. Light grey contours show negative integrated intensity at $-0.9$\,K\,\kms\, and $-1.5$\,K\,\kms. The \dia\, clumps B2-N7 and B2-N8, \amm\, clump B2-A9 (\one) and Class I protostar Elias 33 \citep{elias78} are labelled. b) High resolution $v_{\mbox{\tiny{LSR}}}$ in south-eastern Oph B2. Colour scale is in \kms. c) Fitted $\Delta v$ in \kms. }
\label{fig:mos_l}
\epsscale{1}
\end{figure}

\begin{figure}
\plotone{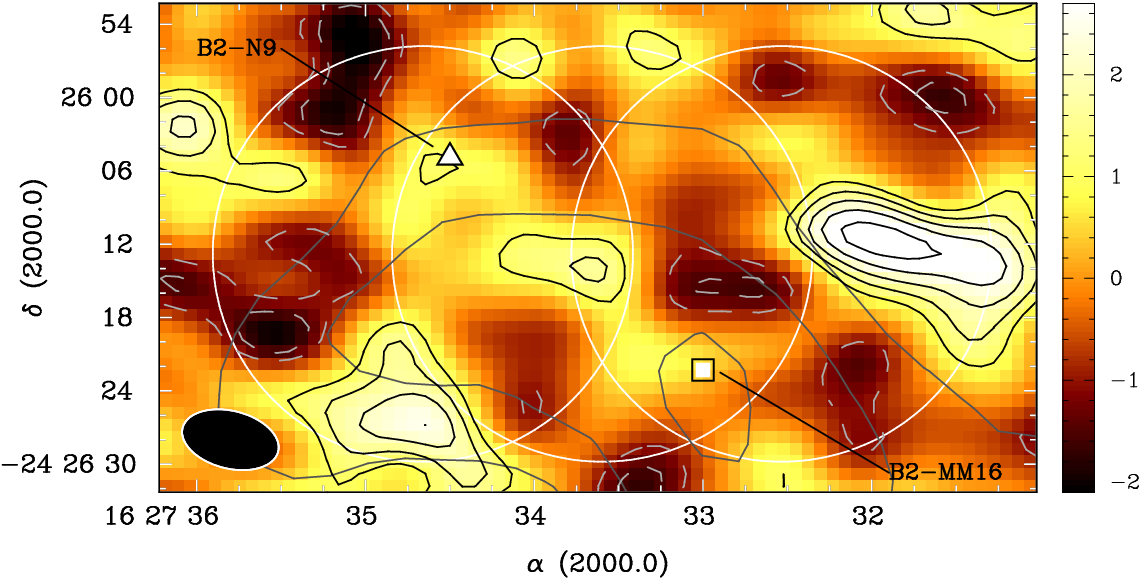}
\caption{\dia\, 1-0 integrated intensity in K\,\kms\, ($T_B$) observed at the ATCA. Black contours begin at 1\,K\,\kms\, and increase by 0.5\,K\,\kms. Light grey contours show negative integrated intensity at $-1$\,K\,\kms\, and $-1.5$\,K\,\kms. White circles show the primary field of view (34\arcsec) of the ATCA at 93\,GHz for each of the three pointings included in this mosaic. Dark grey contours show 850\,\micron\, continuum emission, beginning at 0.3\,Jy\,beam$^{-1}$\, and increasing by 0.1\,Jy\,beam$^{-1}$\,\kms. The \dia\, clump B2-N9 and continuum clump B2-MM16/162733-24262 \citep{motte98,jorgensen08} are labelled. The $\sim 8$\arcsec\, $\times$ 5\arcsec\, FWHM synthesized beam is shown at lower right. }
\label{fig:mos_u}
\end{figure}

In Figures \ref{fig:mm8}a, \ref{fig:nh3peak}a, \ref{fig:mos_l}a and \ref{fig:mos_u}, we show the integrated \dia\, 1-0 intensity towards four of the five regions observed in Oph B2 with the ATCA. In the fifth, most western region observed (see Figure \ref{fig:nobeyama}a), we found no significant interferometer emission above the rms noise. In all plots, the data were integrated over only those velocity channels containing emission from the hyperfine components. \dia, \amm\, and continuum clumps mentioned in the text are labelled on the Figures. 


Figure \ref{fig:mm8}a shows the combined results of three ATCA pointings towards the 850\,\micron\, continuum peak in Oph B2, with the locations of the associated continuum clump B2-MM8 \citep[][also 16273-24271 in \citealt{jorgensen08}]{motte98}, the nearby \dia\, clump B2-N5 identified in \S3.1, and the Class I protostars Elias 32 \citep[][also VSSG17; \citealt{vssg}]{elias78} and Elias 33 (also VSSG18) identified (but note Elias 33 is outside the ATCA primary beam). The single-dish integrated intensity peak, coincident with B2-N5, is offset to the continuum data and B2-MM8 by $\sim 23$\arcsec\, (see Figure \ref{fig:nobeyama}b). Two significant integrated intensity peaks are seen in the ATCA data. The first is coincident with B2-N5 (and consequently offset from B2-MM8), while the second is located between the two protostars where little continuum emission is seen. Some emission is also found to the east and west of B2-MM8. Negative emission features are coincident with the continuum clump peak, and additionally between the two strong \dia\, integrated intensity features. The integrated intensity of the data forms a broken ring-like structure around B2-MM8. The distribution is reminiscent of \dia\, 1-0 emission in B68, where \citet{bergin02} found strong evidence of \dia\, depletion through radiative transfer modelling. Evidence for \dia\, depletion on small scales has also been found in IRAM 04191 \citep{belloche04}. 

In Figure \ref{fig:nh3peak}a, we plot the integrated intensity towards the single pointing in eastern B2, which includes the \dia\, clump B2-N6 and the \amm\, clump B2-A7. \citet{motte98} identify a continuum clump here (B2-MM15, unresolved at 11\arcsec), but no object was identified by \citet{johnstone00} or \citet{jorgensen08} in 850\,\micron\, continuum emission. In the interferometer data, a single integrated intensity peak is seen slightly offset ($\sim 6\arcsec$, significantly less than the Nobeyama beam FWHM of 18\arcsec) NNW from B2-N6. The emission is extended along the beam's long axis, but fitting a 2D Gaussian to the data indicates the emission is also unresolved by the 8\arcsec $\times$ 5\arcsec\, FWHM beam ($\sim 960$\,AU $\times \,600$\,AU at 120\,pc). The peak integrated intensity of the interferometer data is greater relative to the single-dish data in this location by a factor $\sim 2$ compared with the other interferometer observation locations, suggesting that a larger fraction of \dia\, emission is associated with the small-scale peak rather than the extended \dia\, emission.

Figure \ref{fig:mos_l}a shows the results of the south-eastern three pointing mosaic. Strong \amm\, emission, including the \amm\, clump B2-A9, was found in this region in \one. The only continuum emission in the mosaiced area is coincident with the Class I protostar Elias 33. Two overlapping \dia\, clumps identified in \S3.1.1 are shown, with clump peaks separated by $\sim 8$\arcsec. We find two \dia\, integrated intensity peaks in the region, separated by $\sim 30$\arcsec, of which one is coincident with B2-N8. Two regions of negative integrated intensity are found within the primary beam contours. The most negative feature is coincident with Elias 33, while the second is near B2-N7. 

Figure \ref{fig:mos_u} shows the combined results of the three-pointing mosaic in north-eastern Oph B2. One \dia\, clump, B2-N9, is identified in the mosaiced area, which is offset from the continuum clump B2-MM16/16273-24262 by $\sim 30$\arcsec. In the interferometer data, no significant emission is associated with either the \dia\, or continuum clumps. One significant feature is visible in the Figure which is coincident with the western elongation (away from the continuum contours) of the single-dish \dia\, integrated intensity. Multiple negative features are present. 

In general, the small scale structure visible in the integrated intensity maps does not correlate well with objects identified in continuum emission, and is also \edits{sometimes} offset from clumps identified on larger scales from the single-dish \dia\, emission. \edits{The offsets between the single-dish and interferometric \dia\, emission peaks tend to be small, however, and are likely due to the increase in spatial resolution afforded by the interferometer, which is better able to locate precisely the \dia\, emission peak for a given object while resolving out large scale structure.} No small-scale emission was found associated with embedded protostars.  

\section{\dia\, Line Analysis}
\label{sec:analysis}

The \dia\, 1-0 rotational transition has seven hyperfine components \citep{caselli95}. We fit each observed \dia\, 1-0 spectrum with a custom Gaussian hyperfine structure fitting routine written in IDL, described in detail in \one, assuming local thermodynamic equilibrium (LTE) and equal excitation temperatures $T_{ex}$ for each hyperfine component. The expected emission frequencies and emitted line fractions for the hyperfine components were taken from \citet{pagani09}. We fit the spectrum at each pixel and obtain maps of the line-of-sight local standard of rest (LSR) velocity ($v_{\mbox{\tiny{LSR}}}$) of the emission, the line FWHM ($\Delta v$), the total line opacity ($\tau$), and the line brightness ($T_{MB}$). We additionally calculate $T_{ex}$ assuming the line emission fills the beam (the determined $T_{ex}$ will be a lower limit if the observed emission does not entirely fill the beam). 

The line widths determined by the hyperfine structure fitting routine are artificially broadened by the velocity resolution of the observations (0.05\,\kms). We therefore subtracted in quadrature the resolution width, $\Delta v_{res}$, from the observed line width, $\Delta v_{obs}$, such that $\Delta v_{line} = \sqrt{\Delta v_{obs}^2 - \Delta v_{res}^2}$. In the following, we simply use $\Delta v = \Delta v_{line}$ for clarity. 

The uncertainties reported in the returned parameters are those determined by the fitting routine, and do not take the calibration uncertainty ($\sim 8-10$\,\% for the Nobeyama data) into account. The calibration uncertainty does not affect the uncertainties returned for $v_{\mbox{\tiny{LSR}}}$ or $\Delta v$. The excitation temperature, \dia\, column densities, fractional \dia\, abundances and volume densities discussed further below, however, are dependent on the amplitude of the line emission, and are thus affected by the absolute calibration uncertainty.

\begin{figure}
\epsscale{0.6}
\plotone{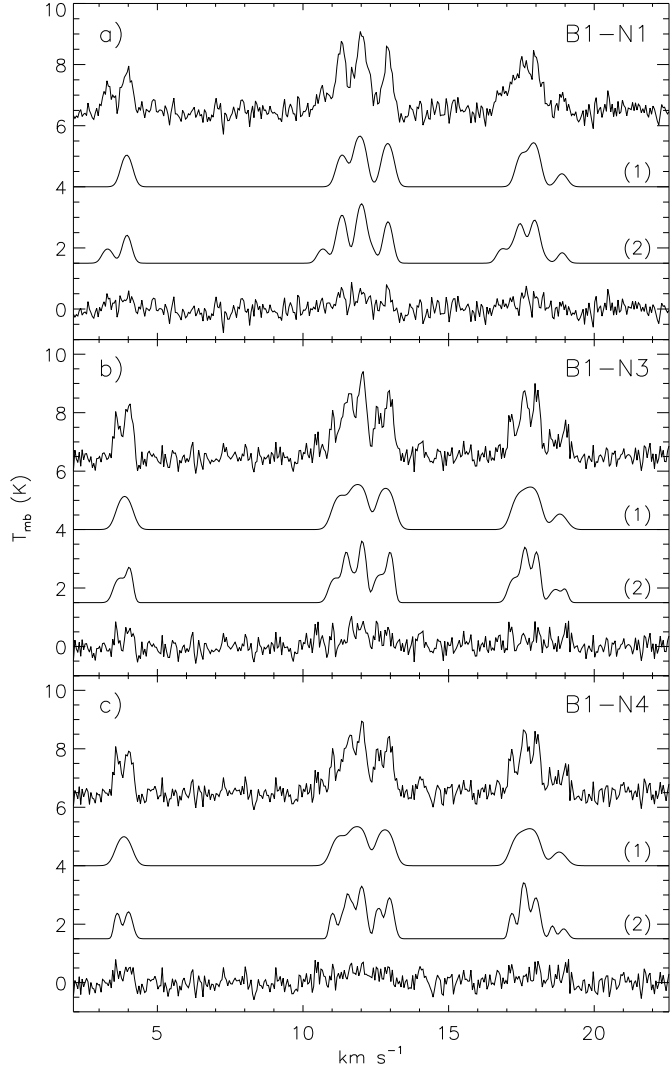}
\caption{Nobeyama \dia\, 1-0 spectra in $T_{MB}$ (K) towards the B1-N1 (a), B1-N3 (b) and B1-N4 (c) \dia\, clump locations, showing results of single and double velocity component HFS fits. For each clump, the best single velocity component (1) and double velocity component (2) fits are shown below the \dia\, 1-0 spectrum, as well as the residual from subtracting the double component fit from the spectrum. For each clump, the spectra and both fits are offset from zero for clarity.}
\label{fig:dblfits}
\epsscale{1}
\end{figure}

Over most of the map, the spectra could be fit with a single set of hyperfine components, but several small regions showed clear double-peaked line profiles in the isolated \dia\, 1-0 hyperfine component, including the \dia\, clumps B1-N1, B1-N3 and B1-N4. To understand better this emission, we fit the spectra with the above Gaussian hyperfine structure fitting routine using two separate velocity components. We show the spectra at the three \dia\, clump peak locations in Figure \ref{fig:dblfits}, overlaid with both single and double velocity component fits. In all cases, we are able to determine well the $v_{\mbox{\tiny{LSR}}}$ and $\Delta v$ of both components, but the fainter component is typically very optically thin and we are unable to constrain simultaneously both $\tau$ and $T_{ex}$. Although the angular separation of B2-N7 and B2-N8 clumps is small, visual inspection of the spectra at the clump peaks did not show a separate velocity component above the rms noise. We discuss the nature of the double-peaked objects further in \S\ref{sec:dblpeak}. In the following sections, we discuss the results of the single velocity component fits only, unless otherwise stated. 


\subsection{Single-Dish Results}

Table \ref{tab:linefit} lists the mean, rms, minimum and maximum values of $v_{\mbox{\tiny{LSR}}}$, $\Delta v$, $\tau$ and $T_{ex}$ found in Oph B. The results across Oph B are shown in Figure \ref{fig:b-fits}, where we plot only data points with $S/N \geqq 2.5$ in the isolated $F_1F \rightarrow F'_1F' = 0 1 \rightarrow 1 2$ component of the \dia\, 1-0 line. In the following sections, we describe in detail the results of the line fitting and examine the resulting line centres and widths, as well as $T_{ex}$ and $\tau$. Table \ref{tab:peak_dat} lists the determined parameters at the peak location of each \dia\, clump. 

\subsubsection{Line Centroids and Widths}
\label{sec:vlsr}

We show in Figures \ref{fig:b-fits}a and \ref{fig:b-fits}b, respectively, the fitted values of $v_{\mbox{\tiny{LSR}}}$ and $\Delta v$ of \dia\, 1-0 across Oph B. Relatively little variation in line-of-sight velocity is seen across Oph B1 and B2, with no evidence of a discontinuity in velocity separating the two sub-Cores. A gradient of increasing $v_{\mbox{\tiny{LSR}}}$ from west to east is seen across Oph B, with the largest $v_{\mbox{\tiny{LSR}}}$ values found in the eastern half of B2, to a maximum $v_{\mbox{\tiny{LSR}}} = 4.44$\,\kms. In Oph B3, the mean $v_{\mbox{\tiny{LSR}}} = 3.06$\,\kms, significantly lower than in B1 ($\langle v_{\mbox{\tiny{LSR}}} \rangle = 3.97$\,\kms) or B2 ($\langle v_{\mbox{\tiny{LSR}}} \rangle = 4.05$\,\kms). The rms variation in $v_{\mbox{\tiny{LSR}}}$ across all Oph B is only 0.23\,\kms, which is dominated by the spread in $v_{\mbox{\tiny{LSR}}}$ in B2 (rms = 0.20\,\kms, compared with 0.12\,\kms\, and 0.14\,\kms\, in B1 and B3, respectively). 

We find greater variation between Oph B1 and B2 in \dia\, 1-0 line width than in $v_{\mbox{\tiny{LSR}}}$. Oph B1 is generally characterized by narrow lines, with a mean $\Delta v = 0.42$\,\kms\, and an rms variation $\sigma_{\Delta v}$ of only 0.08\,\kms. Oph B2 contains regions of both small and large line widths, with a larger mean $\Delta v = 0.68$\,\kms\, and $\sigma_{\Delta v} = 0.21$\,\kms\, than found in B1. Both B1 and B2 have similar minimum line widths ($\Delta v = 0.21$\,\kms\, in B1 and $\Delta v = 0.30$\,\kms\, in B2). Line widths in Oph B3 are extremely narrow, with a mean $\Delta v = 0.27$\,\kms, a minimum $\Delta v = 0.21$\,\kms\, and an rms variation of only $0.04$\,\kms. 

Two locations in B2, both near Class I protostars, show intriguing gradients in line-of-sight velocity and line width, which we show more closely in Figure \ref{fig:b-mm8}. Figure \ref{fig:b-mm8}a shows the first, at a location coincident with the millimeter continuum object B2-MM8, which is found at the thermal dust emission peak in B2 and is the most massive clump identified in Oph B ($M \sim 1\,M_\odot$, \citeauthor{motte98}; \citealt{johnstone00}). As noted previously, MM8 is associated with neither an \dia\, clump nor a maximum of integrated \dia\, 1-0 intensity. The nearby Class I protostar, Elias 32, may be associated with MM8, but is also offset from the clump peak by $\sim 15$\arcsec\, to the southeast. Across the $\sim 30$\arcsec\, width of the clump ($\sim 3600$\,AU at 120\,pc), $v_{\mbox{\tiny{LSR}}}$ varies from $\sim 3.8$\,\kms\, in the west to $\sim 4.2$\,\kms\, in the east. We find no local decrease in line width as might be expected for a starless clump \citep{difran04}. Instead, Figure \ref{fig:b-mm8}b shows how at this location $\Delta v$ increases from east to west, from $\Delta v \sim 0.5$\,\kms\, to $\Delta v \sim 0.9$\,\kms. The resulting velocity gradient, given the $\sim 3600$\,AU diameter of the clump, is $\sim 23$\,\kms\,pc$^{-1}$. The variation in line width in particular appears centered on the nearby protostar. Alternatively, we noted in \one\, a region of wide \amm\, line width located between Oph B2 and B3, which we suggested was due to the overlap of the two Cores with different $v_{\mbox{\tiny{LSR}}}$ along the line of sight. In this case, the wide line widths seen near MM8 also may be due to this overlap. 

Figures \ref{fig:b-mm8}c and \ref{fig:b-mm8}d plot $v_{\mbox{\tiny{LSR}}}$ and $\Delta v$ at the second intriguing gradient, near the Class I protostar Elias 33 in Oph B2. Elias 33 is associated with the millimeter object B2-MM10 \citep[][also 162729-24274 in \citealt{jorgensen08}]{motte98}. Again, no \dia\, clump was found coincident with the protostar. Both $v_{\mbox{\tiny{LSR}}}$ and $\Delta v$ show strong variations at the protostellar location. An abrupt change in $v_{\mbox{\tiny{LSR}}}$ from $\sim 4.1$\,\kms\, to $\sim 3.7$\,\kms\, is seen in over a single 18\arcsec\, beam width ($\sim 2200$\,AU) from west to east. The resulting velocity gradient is $\sim 38$\,\kms\,pc$^{-1}$. A narrow ridge of large line width, $\Delta v \gtrsim 1.0$\,\kms, is found with an elongation perpendicular to the $v_{\mbox{\tiny{LSR}}}$ gradient. On either side, line widths are narrow ($\Delta v \lesssim 0.6$\,\kms). 

\subsubsection{\dia\, Density, Column Density and Fractional Abundance}
\label{sec:nex}

Given the \dia\, 1-0 line characteristics $\Delta v$, $\tau_{tot}$ and $T_{ex}$ determined through HFS fitting, and again assuming LTE, we can calculate the volume density, $n_{ex}$, and the \dia\, column density, $N(\mbox{\dia})$, in Oph B. 

The volume density is calculated following \citet{caselli02} using two-level statistical equilibrium:

\begin{equation}
\frac{n_{ex}}{n^\prime_{cr}} = \frac{T_{ex} - T_{bg}}{(h\nu/k) (1 - T_{ex}/T_{k})}
\label{eqn:nex}
\end{equation}

\noindent where the critical density, corrected for trapping, of the \dia\, 1-0 line is given by: 

\begin{equation}
n^\prime_{cr} = n_{cr} \frac{1 - \exp(-\tau_{\mbox{\tiny{\it hf}}})}{\tau_{\mbox{\tiny{\it hf}}}}.
\end{equation}

\noindent The opacity $\tau_{\mbox{\tiny{\it hf}}}$ is the opacity of a typical hyperfine component. For \dia\, 1-0, we take $\tau_{\mbox{\tiny{\it hf}}} = \tau/9$. We assume that the \dia\, kinetic temperature is equal to the $T_K$ derived in \one\, from \amm\, observations. This assumption introduces some uncertainty in the resulting $n_{ex}$ on the order of $\sim 10 - 50$\,\% if a difference in $T_K$ exists of $\sim 2-5$\,K between gas traced by \amm\, and \dia\, emission (a temperature difference $> 5$\,K is unlikely given the mean $T_K = 15$\,K for Oph B determined in \one\, and the factor $\sim 10-100$ difference in the expected gas densities traced by \amm\, and \dia). Mean, standard deviation, minimum and maximum densities for Oph B1, B2 and B3 are given in Table \ref{tab:col}. For Oph B as a whole, $\langle n_{ex} \rangle = 2.4 \times 10^5$\,\cc, with an rms variation of $2.5 \times 10^5$\,\cc. Separately, the mean $n_{ex}$ for Oph B1 and B2 ($2.0 \times 10^5$\,\cc\, and $3.1 \times 10^5$\,\cc, respectively) are lower by factors of $\sim 2$ and $\sim 3.5$ than the mean volume densities calculated by \citet{motte98} from 1.3\,mm continuum observations. We find only moderate variation in $n_{ex}$ across the Core, and consequently the densities determined for the \dia\, clumps are similar to the mean Core values. This result is discrepant with the $n(\mbox{H$_2$}) \sim 10^7$\,\cc\, densities for the small-scale continuum objects in Oph B, determined by \citeauthor{motte98}, but agrees within a factor of $\sim 2$ with the average 850\,\micron\, clump densities (which we calculated from reported clump masses and radii) from \citet{jorgensen08}. 

\begin{figure}
\plotone{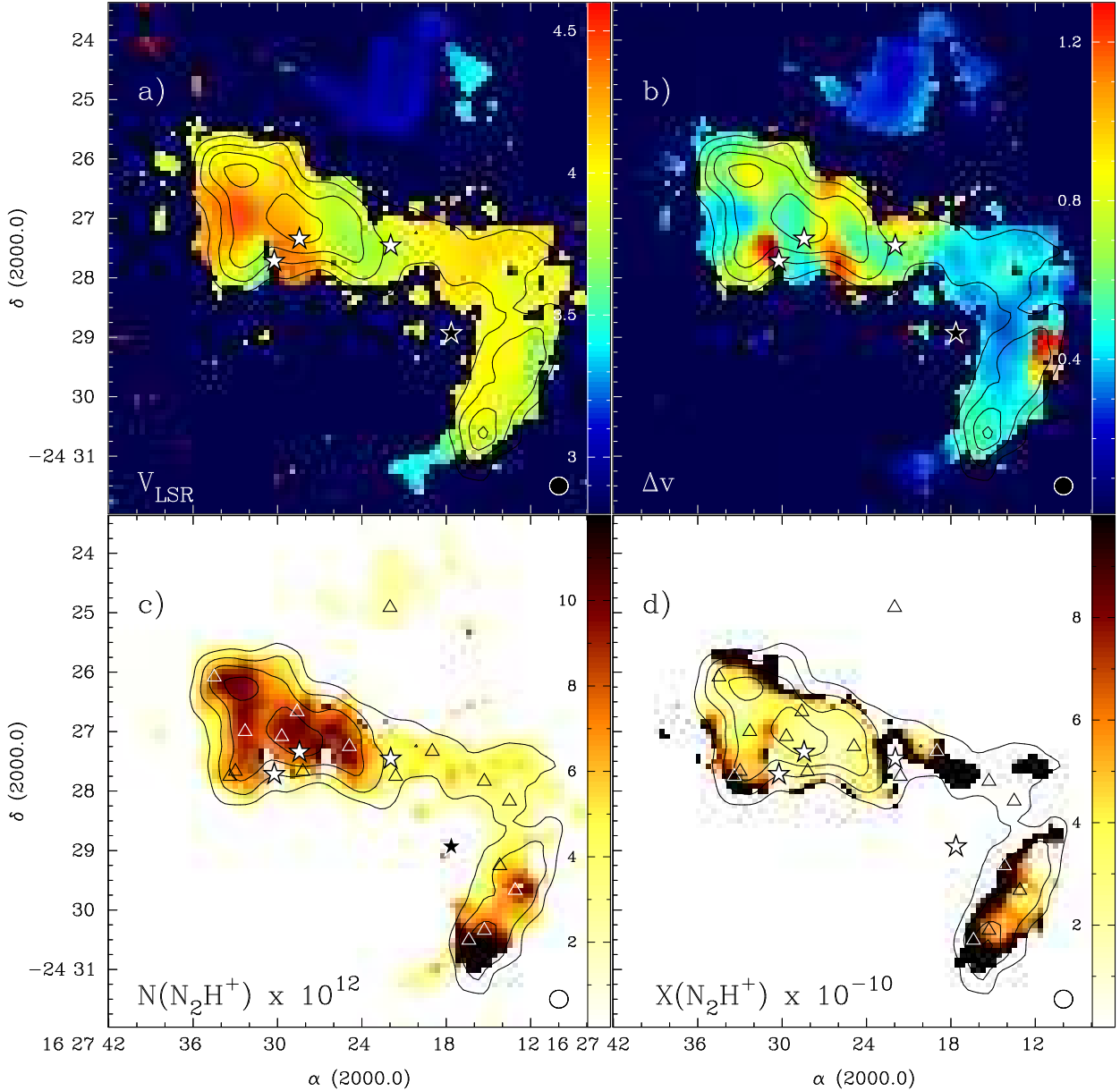}
\caption{a) Line velocity or $v_{\mbox{\tiny{LSR}}}$ in Oph B. Colour
  scale is in \kms. In all plots, contours show integrated \dia\, 1-0
  intensity, beginning at 3.0\,K\,\kms\, and increasing by
  1.5\,K\,\kms. Only pixels where the isolated \dia\, component was
  detected with $S/N \geqq 2.5$ are shown. Stars indicate protostar
  positions. The 18\arcsec\, FWHM beam is shown at lower right. b)
  Fitted $\Delta v$ in Oph B. Colour scale is in \kms. c) \dia\,
  column density $N(\mbox{\dia})$ (cm$^{-2}$) determined using
  Equation \ref{eqn:column}. Values shown have been divided by
  $10^{12}$. The maximum column in Oph B1 is $N(\mbox{\dia}) = 3.3
  \times 10^{12}$\,cm$^{-2}$, but the scale has been truncated to show
  variations in Oph B2. In parts c) and d), triangles indicate peak positions of \dia\,
  clumps discussed in \S\ref{sec:clumps} and listed in Table
  \ref{tab:clump}. d) Fractional abundance $X(\mbox{\dia}) =
  N(\mbox{\dia})\,/\,N(\mbox{H$_2$})$. Only pixels where the
  850\,\micron\, continuum flux density $S_\nu \geqq
  0.1$\,Jy\,beam$^{-1}$ are shown. The maximum fractional abundance $X(\mbox{\dia}) = 4.9 \times 10^{-9}$ is found towards the B1 edge, but the scale has been truncated to show variations in B1 and B2. }
\label{fig:b-fits}
\end{figure}

We follow \citet{difran04} and derive the total \dia\, column density $N(\mbox{\dia})$ using

\begin{equation}
N_{tot} = \frac{3h}{8\pi^3}\frac{1}{\mu^2} \frac{\sqrt{\pi}}{2\sqrt{\ln2}}\frac{1}{(J+1)\exp(-E_J/kT_{ex})}
		\biggl(\frac{kT_{ex}}{hB} + \frac{1}{3}\biggr) \frac{\Delta v \tau}{1 - \exp(-h\nu/kT_{ex})}
\label{eqn:column}
\end{equation}

\noindent where the lower rotational level number $J = 0$, the dipole moment of the \dia\, molecular ion $\mu_e = 3.4$\,D \citep{green74}, the \dia\, rotational constant $B = 46.586702$\,GHz \citep{caselli95} and the rotational level energy above ground of the lower level $E_J = J(J+1)hB$. Here, the rotational partition function $Q_{rot}$ has been approximated by its integral form for a linear molecular ion, $Q_{rot} = kT_{ex}/hB + 1/3$. Note that this calculation is equivalent to that presented by \citet{caselli02ion}. Figure \ref{fig:b-fits}c shows the $N(\mbox{\dia})$ distribution in Oph B. 

\label{sec:nh2}

We also calculate the H$_2$ column density, $N(\mbox{H}_2)$, per pixel in Oph B from 850\,\micron\, continuum data using

\begin{equation}
\label{eqn:dust}
N(\mbox{H}_2) = S_\nu / [ \Omega_m \mu m_{\mbox{{\tiny H}}} \kappa_\nu B_\nu (T_d)] ,
\end{equation}

\noindent where $S_\nu$ is the 850\,\micron\, flux density, $\Omega_m$ is the main-beam solid angle, $\mu = 2.33$ is the mean molecular weight, $m_{\mbox{{\tiny H}}}$ is the mass of hydrogen, $\kappa_\nu$ is the dust opacity per unit mass at 850\,\micron, and $B_\nu (T_d)$ is the Planck function at the dust temperature, $T_d$. As in \one, we take $\kappa_\nu = 0.018$\,cm$^2$\,g$^{-1}$, following \citet{shirley00}, using the OH5 dust model from \citet{ossen94} which describes grains that have coagulated for $10^5$\,years at a density of $10^6$\,\cc\, with accreted ice mantles, and incorporates a gas-to-dust mass ratio of 100. The 15\arcsec\, resolution continuum data were convolved to a final angular resolution of 18\arcsec\, and regridded to match the Nobeyama \dia\, data. 

We assume that the dust temperature $T_d$ per pixel is equal to the kinetic temperature $T_K$ determined in \one\, for \amm\, in Oph B, where we found $\langle T_K \rangle = 15$\,K. This assumption is expected to be good at the densities probed by the \amm\, data ($n \gtrsim 10^4$\cc), when thermal coupling between the gas and dust by collisions is expected to begin \citep{goldsmith78}. Our analysis in \one, however, suggested that \amm\, observations may not trace the densest gas in Oph B. As we expect the denser gas to be colder in the absence of a heating source, this assumption may systematically overestimate $T_d$ in the Core. The derived $N(\mbox{H}_2)$ values would then be systematically low. There is additionally a $\sim 20$\% uncertainty in the continuum flux values, and estimates of $\kappa_\nu$ can vary by $\sim 3$ \citep{shirley00}. Our derived $H_2$ column densities consequently have uncertainties of factors of a few. 

Due to the chopping technique used in the submillimeter observations to remove the bright submillimeter sky, any large scale cloud emission is necessarily removed. As a result, the image reconstruction technique produces negative features around strong emission sources, such as Oph B \citep{johnstone00b}. Emission on scales greater than 130\arcsec\, is suppressed by convolving the original continuum map with a $\sigma = 130$\arcsec\, continuum beam, and subtracting the convolved map from the original data. While the flux density measurements of bright sources are likely accurate, the column densities measured at core edges may be underestimates. Following \one, we thus limit our analysis to pixels where $S_\nu \geq 0.1$\,Jy\,beam$^{-1}$, though the rms noise level of the continuum map is $\sim 0.03$\,Jy\,beam$^{-1}$. For a dust temperature $T_d = 15$\,K, this flux level corresponds to $N(\mbox{H}_2) \sim 6 \times 10^{21}$\,cm$^{-2}$. 

Using $N(\mbox{H$_2$})$ and $N(\mbox{\dia})$, we next calculate per pixel the fractional abundance of \dia\, relative to H$_2$, $X(\mbox{\dia})$, in Oph B. In Figure \ref{fig:b-fits}d, we show the distribution of $X(\mbox{\dia})$ in Oph B. The mean, rms, minimum and maximum values of the derived \dia\, column density and fractional abundance are given in Table \ref{tab:col}, while specific values for individual \dia\, clumps are given in Table \ref{tab:peak_dat}. The $N(\mbox{H$_2$})$ and consequently the $X(\mbox{\dia})$ uncertainties given in Table \ref{tab:peak_dat} include the $\sim 20$\% uncertainty in the submillimeter continuum flux values only; uncertainties in $T_d$ and $\kappa_\nu$ are not taken into account. 

\edits{The mean \dia\, column density $\langle N(\mbox{\dia})\rangle = 5.5 \times 10^{12}$\,cm$^{-2}$ in Oph B. As seen in Table \ref{tab:col}, \dia\, column densities in Oph B2 are slightly greater, on average, than in Oph B1. B1, however, contains the largest \dia\, column density peak, with a maximum $N(\mbox{\dia}) = 1.7 \times 10^{13}$\,cm$^{-2}$ found near (but not at the peak position of) the \dia\, clump B1-N4. The smallest average column density is found towards B3, and the maximum value in B3 is $\sim 2-4$ times less than peak values found in both B1 and B2 ($N(\mbox{\dia}) = 1.1 \times 10^{13}$\,cm$^{-2}$ towards B2-N9; similar high values are also found towards B2-N2). With the exception of the B1 $N(\mbox{\dia})$ peak, observed peaks in $N(\mbox{\dia})$ are only factors of $\sim 2$ above the mean values. }

\edits{In Figure \ref{fig:b-fits}c, we show the locations of \dia\, clumps identified in \S\ref{sec:clumps} over the $N(\mbox{\dia})$ distribution in Oph B. The peak emission locations of \dia\, clumps are often found near, but are not perfectly coincident with, local $N(\mbox{\dia})$ maxima. Clumps identified in continuum emission have varying correspondence with $N(\mbox{\dia})$ maxima. In Oph B1, two of three 850\,\micron\, continuum clumps are coincident with the two largest $N(\mbox{\dia})$ maxima, while in Oph B2 only two of five continuum objects is co-located with an $N(\mbox{\dia})$ maximum. The embedded protostars in B2 are not associated with $N(\mbox{\dia})$ maxima.}

\edits{The mean fractional \dia\, abundance for Oph B, $\langle X(\mbox{\dia})\rangle = 8.0 \times 10^{-10}$, is greater by a factor of $\sim 2 - 4$ than previous measurements in isolated starless cores \citep{caselli02,tafalla02,tafalla06,keto04}. As seen in Table \ref{tab:col}, the mean \dia\, abundance in Oph B1 is greater by a factor $\sim 2$ than in Oph B2, although the maximum values are similar. The greatest \dia\, abundances are found towards the sub-Core edges and the connecting filament. Conversely, the smallest $X(\mbox{\dia})$ values are located at the centres of B1 and B2. Little small-scale variation is found in the sub-Core interiors. We discuss further in \S5.2 the general trend in $X(\mbox{\dia})$ with H$_2$ column density.}

\edits{At the peak locations of \dia\, clumps in B2, the mean $X(\mbox{\dia}) = 3.2 \times 10^{-10}$ is slightly greater than typical values found in isolated starless cores. \dia\, clumps, however, are not located at clear $X(\mbox{\dia})$ maxima at the 18\arcsec\, resolution of the single-dish data, as shown in Figure \ref{fig:b-fits}d where we plot the \dia\, clump locations over the $X(\mbox{\dia})$ distribution in Oph B. In Oph B2, two of five continuum clumps are found towards \dia\, abundance minima, and no clear abundance minima or maxima are associated with the remaining three. In Oph B1, two of three continuum clumps are co-located with $X(\mbox{\dia})$ minima, while a good abundance measurement was not found at the position of the third. } 


\subsection{Interferometer Results}
\label{sec:analysis-atca}

We next look at the line fitting results in the ATCA pointings. Complex line structures were frequently found in the interferometer data (see \S\ref{sec:results-atca}), and in combination with relatively low S/N in many locations a full 7-component Gaussian \dia\, 1-0 line fit did not converge. Instead, we fit a single Gaussian to the isolated \dia\, 1-0 component to determine the line $v_{\mbox{\tiny{LSR}}}$ and $\Delta v$. In Figures \ref{fig:mm8}, \ref{fig:nh3peak} and \ref{fig:mos_l} (parts b and c) we show $v_{\mbox{\tiny{LSR}}}$ and $\Delta v$ over the B2-MM8 region, the eastern \dia\, object B2-N5, and the south-eastern edge of Oph B2, respectively. While emission was found in the north-eastern mosaic (see Figure \ref{fig:mos_u}), the relatively low line strengths compared with the mosaic rms and complex line structures precluded good fits to the data. 

\begin{figure}
\epsscale{0.6}
\plotone{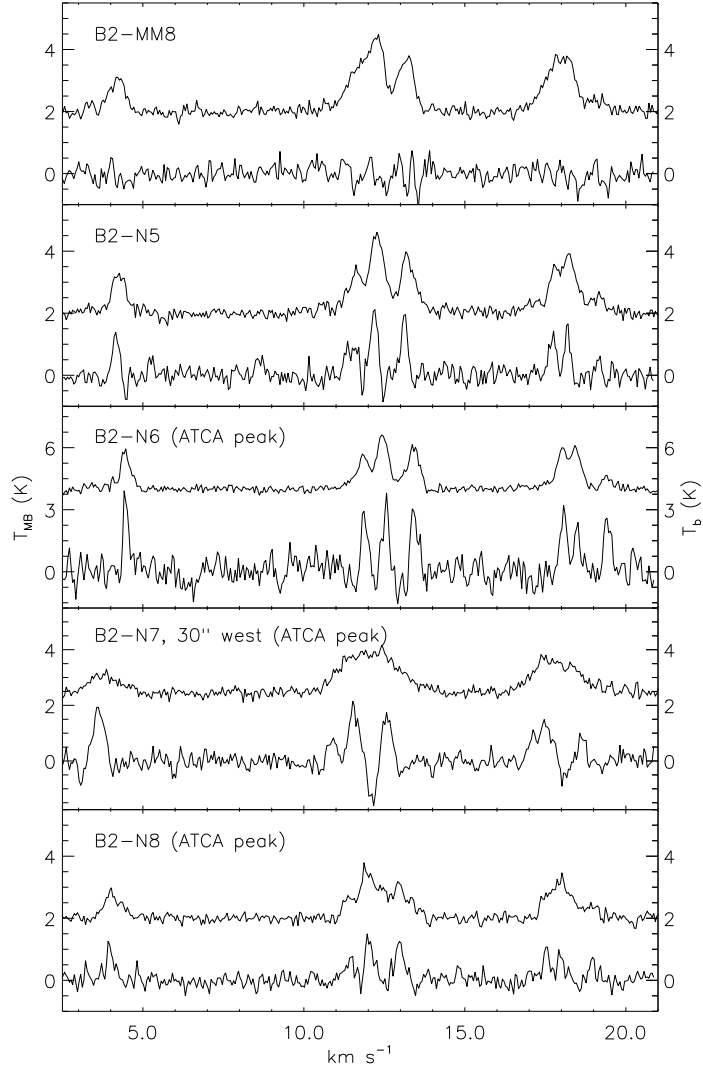}
\caption{\dia\, 1-0 spectra towards B2-MM8, B2-N5, B2-N6, B2-N7 and B2-N8 observed with Nobeyama and the ATCA. Nobeyama spectra have been offset from 0 by 2\,K, 2\,K, 4\,K, 2.5\,K and 2\,K for clarity. The temperature scale is in $T_{MB}$ (K; Nobeyama) and $T_B$ (K; ATCA). Spectra towards B2-N6, B2-N7 and B2-N8 are taken from the ATCA integrated intensity maxima nearby, and are offset from the clumps identified in the single-dish data by $\sim 5 - 30$\arcsec. Note that little interferometer emission is observed towards the MM8 location. ATCA spectra at B2-N5, B2-N6, B2-N7 and B2-N8 have smaller line widths by a factor $\gtrsim 2$ than are observed with the single-dish telescope. The $v_{\mbox{\tiny{LSR}}}$ blueshift of 0.1\,\kms\, observed in the ATCA data relative to the Nobeyama data towards B2-N5 may be a result of negative features in the interferometer data. No offset in velocity was found between the ATCA and Nobeyama observations towards B2-N6, B2-N7 or B2-N8.}
\label{fig:anspecs}
\epsscale{1}
\end{figure}

In general, we find the structures visible in the interferometer data have similar $v_{\mbox{\tiny{LSR}}}$ within uncertainties, and universally narrower $\Delta v$ by a factor $\sim 2$, compared with the single-dish data at the same location. This difference can be seen qualitatively in Figure \ref{fig:anspecs}, where we plot the Nobeyama and ATCA spectra towards B2-MM8, B2-N5, B2-N6, B2-N7 and B2-N8. (We show the B2-N6, B2-N7 and B2-N8 spectra at the interferometer data peak, as the ATCA data show integrated intensity maxima offset by $\sim 5$\arcsec\, from the single-dish data towards B2-N6 and B2-N8, while significant interferometer emission was not seen towards B2-N7.) The B2-N5 $v_{\mbox{\tiny{LSR}}}$ is blueshifted by 0.1\,\kms\, in the ATCA data relative to the Nobeyama data, but Figure \ref{fig:anspecs} shows this is likely due to strong negative features in the interferometer spectrum. 

We find a velocity gradient of $\sim 0.2$\,\kms\, over $\sim 10$\arcsec\, (0.005\,pc, or $\sim 33$\,\kms\,pc$^{-1}$) with a position angle of $\sim 75$\degr\, across the small interferometer peak north of B2-N6 (see Figure \ref{fig:nh3peak}b). If this gradient is associated with rotation of the clump, this value is large compared with typical gradient values observed at low resolution, but is comparable to high resolution \dia\, 1-0 results for dense starless and protostellar clumps. In Ophiuchus A, \citet{difran04} found velocity gradients of 0.5\,\kms\, over 0.01\,pc, or 50\,\kms\,pc$^{-1}$, towards several submillimeter clumps. In addition, \citet{lee07} found a velocity gradient of $30\,/\,\cos (i)$\,\kms\,pc$^{-1}$) within 30\arcsec\, (0.04\,pc at 300\,pc)  towards the L1521B protostellar group, perpendicular to an observed outflow direction and suggestive of fast rotation. The gradient seen in L1521B is both an order of magnitude larger than that seen at lower resolution and in the opposite direction (a neighbouring core was included in the low resolution data). In the protostellar core IRAM 04191, \citet{belloche04} found a mean velocity gradient of $\sim 17$\,\kms\,pc$^{-1}$ through interferometer \dia\, observations, which is a factor $\sim 5$ larger than that seen in single-dish data. 

B2-N6 is also associated in the interferometer data with extremely narrow \dia\, 1-0 line widths, with a mean $\Delta v = 0.21 \pm 0.04$\,\kms\, averaged over an 8\arcsec\, $\times$ 5\arcsec\, area matching the beam FWHM. The observed single-dish line width at the same location is $0.39\pm0.01$\,\kms, listed in Table \ref{tab:peak_dat}. These are the narrowest lines found in the Oph B2 Core. (Slightly greater line widths are found in Oph B1 towards the B1-N3 and B1-N4 clumps when fitted by a two-velocity component structure in the single-dish data, see \S\ref{sec:analysis}.) The velocity gradient, small size and narrow line widths associated with B2-N6 in the interferometer data make it an interesting \edits{emission feature} in B2, which we discuss further in \S\ref{sec:n6}.

\section{Discussion}

\subsection{General trends}
\label{sec:trends}

\begin{figure}
\plotone{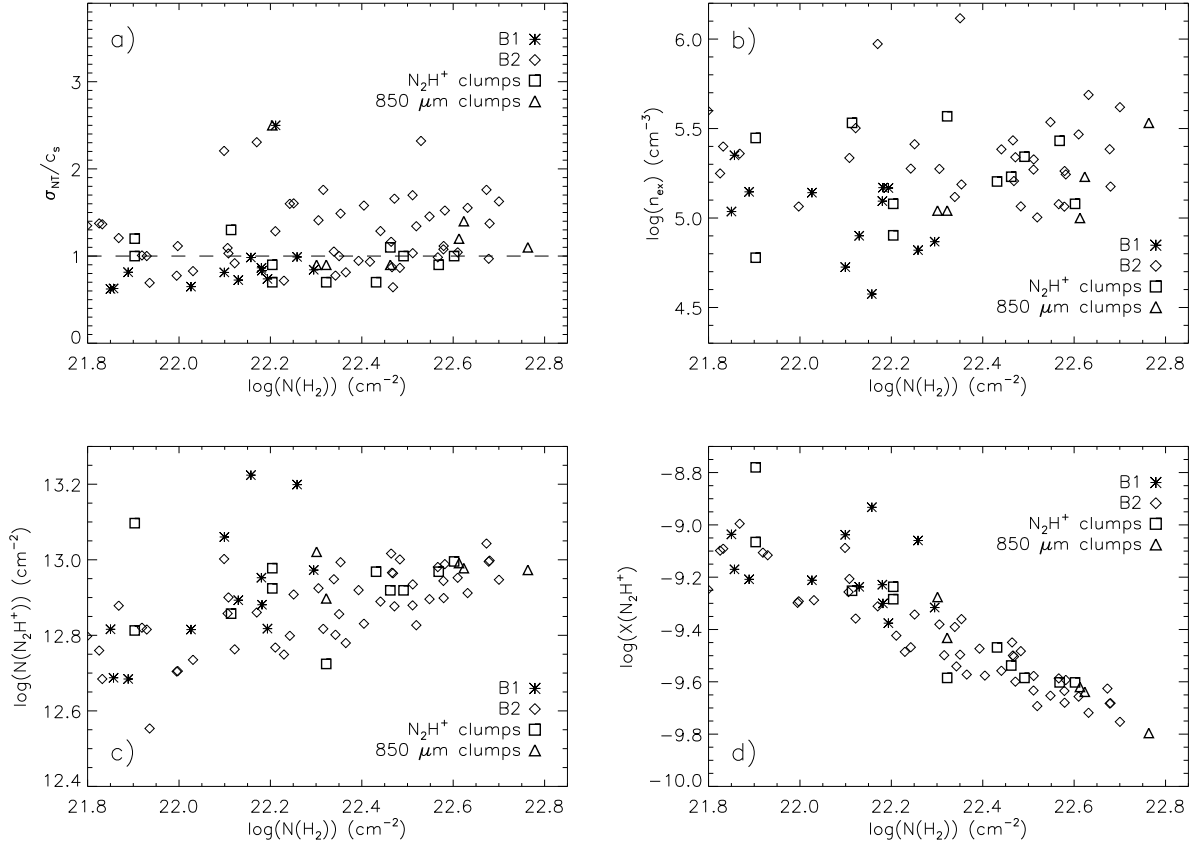}
\caption{a) Ratio of the non-thermal line width to the sound speed, $\sigma_{\mbox{\tiny{NT}}}\,/\,c_s$, versus $N(\mbox{H$_2$})$ (cm$^{-2}$, derived from 850\,\micron\, continuum data) in Oph B1 and B2. Each point represents an 18\arcsec\, pixel (matching the beam FWHM). The dashed line indicates $\sigma_{\mbox{\tiny{NT}}}\,/\,c_s = 1$. Most \dia\, clumps have subsonic non-thermal motions, as does a large fraction of the \dia\, emission in both Oph B1 and B2. Note that the velocity resolution of the data, $\Delta v_{res} = 0.05$\,\kms, is significantly smaller than the apparent cutoff in the non-thermal to sound speed ratio at $\sigma_{\mbox{\tiny{NT}}}\,/\,c_s \sim 0.5$. In all plots, also shown are values at the peak positions of \dia\, clumps identified with {\sc clumpfind} as described in \S3 and 850\,\micron\, clump locations identified by \citet{jorgensen08}. b) Distribution of $n_{\mbox{\tiny{ex}}}$ (cm$^{-3}$) with $N(\mbox{H$_2$})$ in Oph B1 and B2.  c) $N(\mbox{\dia})$ (cm$^{-2}$) versus $N(\mbox{H$_2$})$ in Oph B1 and B2. There is no clear trend in $N(\mbox{\dia})$ seen in Oph B1, but in Oph B2 $N(\mbox{\dia})$ appears to increase with $N(\mbox{H$_2$})$. d) $X(\mbox{\dia})$ versus $N(\mbox{H$_2$})$ in Oph B1 and B2.} 
\label{fig:n2h_trends}
\end{figure}

We show in Figure \ref{fig:n2h_trends} the distribution of the ratio of the non-thermal dispersion to the sound speed, $\sigma_{\mbox{\tiny{NT}}}\,/\,c_s$, $n_{ex}$, $N(\mbox{\dia})$ and $X(\mbox{\dia})$ with $N(\mbox{H$_2$})$ in Oph B, omitting pixels where the 850\,\micron\, continuum flux $S_\nu  \leq 0.1$\,Jy\,beam$^{-1}$. Data points represent values for 18\arcsec\, pixels, i.e., approximately the Nobeyama beam FWHM. We show individually pixels in Oph B1 and B2, and also show peak values for \dia\, clumps, identified in \S\ref{sec:clumps}, and 850\,\micron\, continuum clumps \citep{jorgensen08}. Note that no Oph B3 values are plotted due to the lack of continuum emission at the B3 location. The Figure shows that the \dia\, clumps do not reside at the highest H$_2$ column densities, but scatter over the range of $N(\mbox{H$_2$})$ calculated above our submillimeter flux threshold. The submillimeter clumps tend to be found at higher $N(\mbox{H$_2$})$ values than the \dia\, clumps, on average, but also show a spread in peak $N(\mbox{H$_2$})$. 

In \one, we used the ratio of \amm\, (1,1) and (2,2) emission lines in Oph B to determine the kinetic temperature $T_K$ in each pixel, which we can use to calculate $\sigma_{\mbox{\tiny{NT}}}$ and $c_s$ for \dia\, emission in Oph B. Given $T_K$, $\sigma_{\mbox{\tiny{NT}}} = \sqrt{\sigma_{obs}^2 - k_B T_K\,/\,(\mu_{mol} m_{\mbox{\tiny{H}}})}$, where $k_B$ is the Boltzmann constant, $m_{\mbox{\tiny{H}}}$ is the mass of the hydrogen atom, $\mu_{mol}$ is the molecular weight in atomic units ($\mu_{\mbox{\tiny{\dia}}} = 29.02$) and $\sigma_{obs} = \Delta v\,/\,(2\sqrt{2\ln2})$. We list values and propagated uncertainties for individual \dia\, clumps in Table \ref{tab:peak_dat}, and give the mean, rms, minimum and maximum $\sigma_{\mbox{\tiny{NT}}}\,/\,c_s$ in Table \ref{tab:col} for each of Oph B1, B2 and B3. 

We find a mean $\sigma_{\mbox{\tiny{NT}}}\,/\,c_s = 1.02$ for \dia\, 1-0 emission across Oph B with an rms variation of 0.49, indicating that the non-thermal motions are approximately equal, on average, to the sound speed. When looked at separately, we find a difference in the  $\sigma_{\mbox{\tiny{NT}}}\,/\,c_s$ ratio between Oph B1 and B2. In Oph B1, the mean $\sigma_{\mbox{\tiny{NT}}}\,/\,c_s = 0.71$, with an rms variation of 0.16. In Oph B2, the mean $\sigma_{\mbox{\tiny{NT}}}\,/\,c_s = 1.26$ with an rms variation of 0.4. Oph B3 is dominated by nearly thermal line widths, with the mean $\sigma_{\mbox{\tiny{NT}}}\,/\,c_s = 0.39$ and an rms variation of only 0.13. 

We find little variation in $\sigma_{\mbox{\tiny{NT}}}\,/\,c_s$ with $N(\mbox{H$_2$})$ in Oph B, as shown in Figure \ref{fig:n2h_trends}a. The difference in the mean $\sigma_{\mbox{\tiny{NT}}}\,/\,c_s$ ratio between B1 and B2 is clear, as is the significantly larger scatter of non-thermal line widths in B2 than in B1. The scatter in $\sigma_{\mbox{\tiny{NT}}}\,/\,c_s$ extends to the highest H$_2$ column densities. In B1, all non-thermal line widths $\sigma_{\mbox{\tiny{NT}}} \leqq c_s$, shown by the dashed line. Similar results are found for most \dia\, clumps, and approximately half the continuum clumps. At 18\arcsec\, resolution, the observed non-thermal line widths do not extend to arbitrarily low values, as the data show a cutoff in the $\sigma_{\mbox{\tiny{NT}}}\,/\,c_s$ ratio at $\sigma_{\mbox{\tiny{NT}}}\,/\,c_s \sim 0.5$ for both B1 and B2, below which no data points are found (but note that in \S\ref{sec:analysis-atca}, we found narrow line widths at high resolution). For $T_K = 15$\,K, the mean temperature found in Oph B, the $\sigma_{\mbox{\tiny{NT}}}\,/\,c_s$ cutoff corresponds to $\sigma_{\mbox{\tiny{NT}}} \sim 0.12$\,\kms. Note that the velocity resolution of the \dia\, data, $\Delta v_{res} = 0.05$\,\kms, is significantly less than the observed lower $\sigma_{\mbox{\tiny{NT}}}\,/\,c_s$ limit. 

In \one, we discussed the possible origins of the large non-thermal motions in Oph B, and based on timescale arguments ruled out primordial motions (i.e., motions inherited from the parent cloud). It is interesting that most of the gas traced by \dia\, 1-0 in Oph B2 remains dominated by transonic non-thermal motions, while the line widths in Oph B1 reveal the Core is significantly more quiescent at the $\sim 10^5$\,\cc\, densities traced by \dia\, 1-0 emission. The most obvious source of this difference is the presence of embedded protostars in Oph B2 while Oph B1 is starless. \citet{kamazaki03} found an outflow in CO 3-2 emission in Oph B2 centred near Elias 33 and Elias 32 and oriented approximately east-west, but were unable to pinpoint which protostar was the driving source. In upcoming results from the JCMT Gould Belt Legacy survey \citep{gbls}, which mapped Oph B in CO 3-2, $^{13}$CO 3-2 and C$^{18}$O 3-2 at 14\arcsec\, resolution, this highly clumped outflow is shown to extend over more than 10\arcmin\, (i.e., 0.4\,pc at 120\,pc, White {\it et al.} 2009, in preparation). The outflow axis is also aligned such that the outflow does not appear to be impacting Oph B1 significantly, potentially explaining the difference in non-thermal line widths between the two sub-Cores\footnotemark\footnotetext{JCMT Spring 2009 newsletter, http://www.jach.hawaii.edu/JCMT/publications/newsletter/n30/jcmt-n30.pdf}. 


\dia\, is not generally thought to be a tracer of protostellar outflows, since it is expected to be destroyed quickly through reactions with CO, which evaporates from dust grains in the higher temperature gas near the driving protostar. On small scales in B2, however, some variations in $v_{\mbox{\tiny{LSR}}}$ and $\Delta v$ appear correlated with the presence of nearby Class I protostars (see \S\ref{sec:vlsr} and Figure \ref{fig:b-mm8}). \citet{chen08} suggest \dia\, can be entrained in protostellar jets before a molecular outflow releases CO from grain surfaces which then can destroy \dia. In a small \dia\, 1-0 survey of the Serpens NW cluster, \citet{williams00} found that the largest $\sigma_{\mbox{\tiny{NT}}}$ values ($\sigma_{\mbox{\tiny{NT}}} > 0.6$\,\kms, greater than seen here in Oph B2) occurred in cores containing protostellar sources which power strong outflows. If caused by the substantial protostellar outflow, the large line widths on a global scale in Oph B2 suggest that protostellar outflows are able to inject additional turbulence into the {\it high density} gas in cluster-forming Cores, thereby increasing the mass at which clumps become gravitationally unstable and altering the consequent fragmentation and evolution of existing embedded clumps. \citet{difran01} did not see this effect in the \dia\, 1-0 observations of the circumstellar dense gas associated with the protostellar, outflow-driving source NGC 1333 IRAS4B, suggesting that the ability of the outflow to inject turbulent energy may be determined by additional factors, such as the collimation or strength of the outflow. A comparison of the outflow properties in these regions would be useful to probe further the impact of outflows on cluster forming dense gas. 

Alternatively, the larger line widths in B2 may be due to global infall motions. In \one, we determined a virial mass $M_{vir} \sim 8$\,M$_\odot$ for Oph B2 based on \amm\, line widths, which is a factor $\sim 5$ less than Core mass estimates based on thermal dust continuum emission (with a factor $\sim 2$ uncertainty). In contrast, $M/M_{vir} \sim 1$ in Oph B1. Some evidence for infall has been observed in self-absorbed molecular line tracers in Oph B2 \citep{andre07,gurney08}, but it is difficult to determine unambiguously given the complex outflow motions also present. Additionally, the consistent line widths found in \amm\, gas for both B1 and B2 (discussed further in \S\ref{sec:sig} in comparison with the \dia\, results) suggest a common source for non-thermal motions at gas densities $n \sim 10^{3-4}$\,\cc, which then impacts differently the gas at higher densities in the two sub-Cores. 



Figure \ref{fig:n2h_trends}b shows the volume density $n_{\mbox{\tiny{ex}}}$ as a function of $N(\mbox{H$_2$})$. Nearly all pixels have $n_{\mbox{\tiny{ex}}} \gtrsim 10^5$\,\cc, and we find slightly greater $n_{\mbox{\tiny{ex}}}$ at higher column densities. The \dia\, and continuum clumps scatter over the full range of density values found in Oph B. 

We find a slight trend of increasing \dia\, column density, $N(\mbox{\dia})$, with $N(\mbox{H$_2$})$, shown in Figure \ref{fig:n2h_trends}c, although the mean $N(\mbox{\dia})$ changes by less than a factor of $\sim 2$ over the order of magnitude range in $N(\mbox{H$_2$})$ found in Oph B. Several pixels in Oph B1, and one \dia\, clump, have significantly greater $N(\mbox{\dia})$ for their $N(\mbox{H$_2$})$ values compared with the rest of the data. The small gradient in $N(\mbox{\dia})$ with $N(\mbox{H$_2$})$ leads to a trend of {\it decreasing} fractional \dia\, abundance with increasing H$_2$ column density, shown in Figure \ref{fig:n2h_trends}d. We find \dia\, abundances decrease by an order of magnitude between the limiting $N(\mbox{H$_2$})$ threshold and the peak $N(\mbox{H$_2$})$ values. We fit a linear relationship to $\log X(\mbox{\dia})$ versus $\log N(\mbox{H$_2$})$ for Oph B1 and B2 separately. In B1 alone, where we have relatively few data points, we find $\log X(\mbox{\dia})$ is consistent within the uncertainties with being constant with $N(\mbox{H$_2$})$, but in conjunction with the Oph B2 data, we find

\begin{equation}
\label{eqn:xvsh}
\log X(\mbox{\dia}) = (7.1 \pm 0.9) - (0.74 \pm 0.04) \times \log N(\mbox{H$_2$})
\end{equation}

\noindent using the same H$_2$ column density limits as in Figure \ref{fig:n2h_trends}. 

\subsection{Small-Scale Features}

We next discuss the physical properties of the small-scale structure present in Oph B, including the \dia\, clumps, continuum clumps, and embedded protostars, and describe in detail the compact, thermal object B2-N6.

\subsubsection{Comparison of \dia\, clumps, continuum clumps and protostars}
\label{sec:compare_clumps}

We list in Table \ref{tab:prot_dat} the physical parameters derived from \dia\, 1-0 emission towards 850\,\micron\, continuum clump locations \citep{jorgensen08} and embedded Class I protostars \citep{enoch09}. Columns are the same as in Table \ref{tab:peak_dat}. In Table \ref{tab:mean_dat}, we list the mean values of each physical parameter towards \dia\, clumps, continuum clumps and protostars. Note that we were only able to solve for all parameters listed towards a single protostar, and discuss below only those mean parameters where we obtained values from three or more objects. 

We find no difference in the mean $v_{LSR}$ between \dia\, clumps, continuum clumps and protostars. There is a clear reduction in $\Delta v$ towards \dia\, clumps relative to continuum clumps (mean $\Delta v = 0.49$\,\kms and 0.68\,\kms, respectively), but excluding one continuum clump (162712-24290) reduces the mean $\Delta v$ to 0.58\,\kms. Protostars are associated with wider mean $\Delta v$, but the larger mean is also driven by large $\Delta v$ towards a single object. The differences between \dia\, and continuum clumps are marginally significant given the variance of $\Delta v$ across the entire Core is only 0.08\,\kms\, in Oph B1 and 0.21\,\kms\, in Oph B2. We find similar \dia\, line opacities and excitation temperatures towards continuum and \dia\, clumps. Continuum clumps are associated with larger non-thermal motions, with a mean $\sigma_{NT} / c_s = 1.24$ which is 1.5 times that associated with \dia\, clumps (mean $\sigma_{NT}/c_s = 0.86$). \dia\, column densities are similar towards \dia\, and continuum clumps (within $\sim 25$\,\%), but \dia\, clumps are associated with smaller $N(\mbox{H$_2$})$ and hence greater $X(\mbox{\dia})$ by a factor $> 2$. In fact, the mean \dia\, abundance towards continuum clumps is less than $X(\mbox{\dia})$ averaged over the entire Oph B Core. Derived volume densities are the same for \dia\, and continuum clumps. 

In summary, the \dia\, clumps are associated with smaller line widths and subsonic non-thermal motions relative to continuum clumps and protostars, but the difference in mean values is similar to the variance in line widths across the Core. The most significant difference between \dia\, and continuum clumps are found in their \dia\, abundances, with a mean $X(\mbox{\dia})$ towards continuum clumps that is lower than $X(\mbox{\dia})$ in \dia\, clumps. 

\subsubsection{Oph B2-N6}
\label{sec:n6}

The B2-N6 clump is a pocket of narrow \dia\, 1-0 line emission within the larger, more turbulent B2 Core, with a large fraction ($\gtrsim 50$\%) of the emission at the clump peak coming from the small size scales probed by the ATCA observations (see Figure \ref{fig:nh3peak}). Although we do not consider all \dia\, clumps to trace underlying physical structure in Oph B, the remarkable properties of B2-N6 suggest that further attention be paid to that particular clump. Line widths measured from the interferometer data are a factor $\sim 2$ less than those recorded from the single-dish data ($\Delta v = 0.21 \pm 0.04$\,\kms\, compared with $0.39 \pm 0.01$\,\kms). The millimeter continuum object B2-MM15, identified through a spatial filtering technique \citep{motte98}, lies offset by $\sim 12$\arcsec\, from the peak interferometer integrated intensity (see Figure \ref{fig:nh3peak}).  \citeauthor{motte98} calculate a mass of only $0.17$\,M$_\odot$ associated with the B2-MM15 clump, but based on its small size (unresolved at 15\arcsec\, resolution, using an estimate of $r \sim 800$\,AU) derive an average density of $n \sim 1.2 \times 10^8$\,\cc. This density is significantly larger than the densities calculated in \S\ref{sec:nex} at 18\arcsec\, resolution, but higher sensitivity \dia\, 1-0 interferometer observations are needed to probe the clump density on small scales. 

We calculated in \one\, a gas kinetic temperature $T_K = 13.9\pm1.1$\,K for the co-located B2-A7 \amm\, (1,1) clump. We find $\sigma_{\mbox{\tiny{NT}}} = 0.06$\,\kms\, for B2-N6, equal to the expected \dia\, thermal dispersion, and $\sigma_{\mbox{\tiny{NT}}}/c_s = 0.26$. A similar compact \dia\, 1-0 clump with nearly thermal line widths was found in Oph A \citep[Oph A-N6, ][]{difran04}. While both clumps are associated with larger-scale \dia\, emission, these objects are only distinguishable as compact, nearly purely thermal clumps when observed with interferometric resolution and spatial filtering. 

\citeauthor{difran04} proposed Oph A-N6 may be a candidate thermally dominated, critical Bonnor-Ebert sphere embedded within the more turbulent Oph A Core. Such objects have sizes comparable to the cutoff wavelength for MHD waves and are in equilibrium between their self-gravity and internal and external pressures.  \citet{myers98} called these objects `kernels', and showed that they could exist in cluster-forming cores with FWHM line widths $\Delta v > 0.9$\,\kms\, and column densities $N(\mbox{H$_2$}) > 10^{22}$\,\cc, with sizes comparable to the spacing of protostars in embedded clusters, suggesting that a population of kernels within a dense Core could form a stellar cluster. The physical conditions in both Oph A and Oph B2 fit these requirements. The B2-N6 clump, however, is significantly smaller (0.004\,pc compared with a predicted $\sim 0.03$\,pc) and less massive (0.17\,M$_\odot$ compared with a predicted $\sim 1$\,M$_\odot$) than the objects discussed by \citeauthor{myers98}. 

If we calculate the mass of a critical Bonnor-Ebert sphere given the ATCA \dia\, 1-0 line width and a radius of $\sim 800$\,AU, where $M/R = 2.4 \sigma^2/G$ \citep{bonnor56}, we find $M_{\mbox{\tiny{BE}}} = 0.02$\,M$_\odot$, or approximately $10\times$ smaller than the mass determined by \citeauthor{motte98} We find a virial mass $M_{vir} = 0.04$\,M$_\odot$, assuming B2-N6 is a uniform sphere and $M_{vir} = 5 \sigma^2 R/G$. If the clump mass is accurate, this suggests that while small, B2-N6 may be gravitationally unstable. \citet{andre07} made a tentative detection of infall motions towards the clump through observations of optically thick lines such as CS, H$_2$CO or HCO$^+$. Thus B2-N6 may be at a very early stage of clump formation, and may eventually form an additional low mass protostar to the three YSOs already associated with dense gas in Oph B2. 


\subsection{Comparison of \dia\, and \amm\, emission in Oph B}

We next compare the physical properties of the gas in Oph B derived here from \dia\, observations with those derived from \amm\, observations, described in \one. In the following discussion, all comparisons are made after convolving the \amm\, data to a final angular resolution of 18\arcsec\, to match the Nobeyama \dia\, observations.

\subsubsection{$v_{\mbox{\tiny{LSR}}}$ and $\sigma_{\mbox{\tiny{NT}}}\,/\,c_s$}
\label{sec:sig}

\begin{figure}
\begin{center}
\plotone{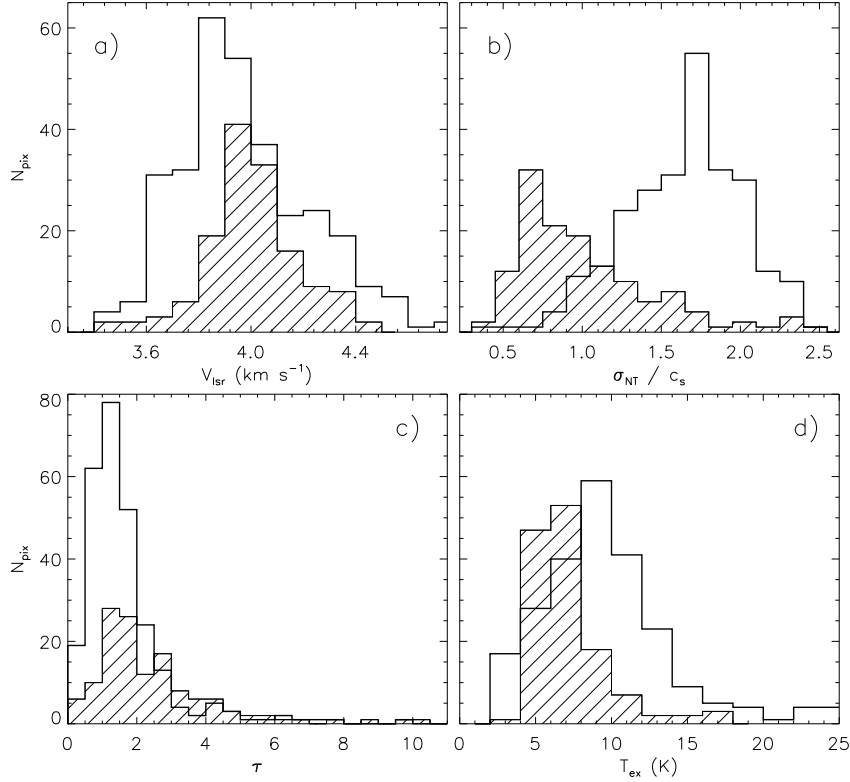}
\caption{a) Comparison of  $v_{\mbox{\tiny{LSR}}}$ from \dia\, 1-0 (hashed) and \amm\, (1,1) (non-hashed) determined from hyperfine structure fitting in Oph B. No significant difference is seen in the distribution of line-of-sight velocities between the two dense gas tracers. Only data points in Oph B1 and B2 are included, since B3 was not detected in \amm\, emission in \one\, with sufficient significance to perform HFS fitting. b) Comparison of $\sigma_{\mbox{\tiny{NT}}}\,/\,c_s$ from \dia\, 1-0 (hashed) and \amm\, (1,1) (non-hashed)  determined from hyperfine structure fitting. The thermal sound speed $c_s$ is calculated using $T_K$ values determined from the ratio of \amm\, (1,1) and (2,2) amplitudes in \one. Non-thermal motions in gas traced by \dia\, are significantly smaller than in gas traced by \amm. c) Comparison of line opacities $\tau$ from \dia\, 1-0 (hashed) and \amm\, (1,1) (non-hashed) summed over all hyperfine components. d) Comparison of $T_{ex}$ \dia\, 1-0 (hashed) and \amm\, (1,1) (non-hashed) determined from hyperfine structure fitting. The \amm\, $T_{ex}$ values have a larger spread and higher mean value ($\langle T_{ex}\rangle = 10$\,K) than \dia\, ($\langle T_{ex}\rangle = 7$\,K).}
\label{fig:compare_nh3}
\end{center}
\end{figure}

We first look at the velocities and non-thermal line widths of the gas traced by \dia\, gas and those determined from \amm\, emission in \one.  Figure \ref{fig:compare_nh3}a shows the distribution of $v_{\mbox{\tiny{LSR}}}$ derived from \dia\, and \amm\, emission in Oph B. There is no significant difference in the measured line-of-sight velocity between the two dense gas tracers, with the mean $v_{\mbox{\tiny{LSR}}} = 3.98$\,\kms\, for \amm\, and 4.01\,\kms\, for \dia. 

In Figure \ref{fig:compare_nh3}b, we show the distribution of $\sigma_{\mbox{\tiny{NT}}}\,/\,c_s$ as determined with \dia\, or \amm\, using the \amm-derived $T_K$ values. The non-thermal line widths measured in \dia\, 1-0 are substantially smaller than those measured in \amm\, in \one, and additionally the variation in the relative magnitude of the non-thermal motions between Oph B1 and B2 was not found in \amm\, emission. The mean \amm\, $\sigma_{\mbox{\tiny{NT}}}\,/\,c_s = 1.64$ in Oph B, and does not vary significantly between B1 ($\langle \sigma_{\mbox{\tiny{NT}}}\,/\,c_s\rangle = 1.62$) and B2 ($\langle \sigma_{\mbox{\tiny{NT}}}\,/\,c_s\rangle = 1.68$). 


The assumption that $T_K$ is similar for both \dia\, and \amm\, is reasonable if both molecules are excited in the same material. The good agreements between the locations of \amm\, and \dia\, clumps, illustrated in Figure \ref{fig:b_sep}, and the $v_{\mbox{\tiny{LSR}}}$, illustrated in Figure \ref{fig:compare_nh3}a, support this assumption. The substantial offset in the $\sigma_{\mbox{\tiny{NT}}}\,/\,c_s$ ratio seen in Figure \ref{fig:compare_nh3}b, however, between \amm\, and \dia\, emission indicates significantly different motions are present in the gas traced by \amm\, than in the gas traced by \dia. The gas densities traced by \dia\, are $\sim 1-2$ orders of magnitude greater than those traced by \amm\, ($\sim 10^5$\,\cc\, and $\sim 10^{3-4}$\,\cc, respectively). In fact, if we attempt to derive the gas kinetic temperature by assuming equal non-thermal motions for the \dia\, 1-0 and \amm\, (1,1) emission (effectively assuming \dia\, and \amm\, trace the same material), the resulting average value is a highly unlikely $T_K = 190$\,K for Oph B2. Given the narrow \dia\, lines observed in Oph B1, no physical solution for $T_K$ can be found under the assumption of equal $\sigma_{\mbox{\tiny{NT}}}$. 

Starless cores are typically found to be well described by gas temperatures that are either constant or decreasing as a function of increasing density \citep{difran07}. Given that \dia\, traces denser gas, we then expect $T_K$ from \amm\, measurements to be biased high in starless cores, e.g., gas traced by \dia\, should be colder than gas traced by \amm. The mean $T_K$ found in Oph B in \one\, was 15\,K. With a lower $T_K$, the sound speed would be smaller and the returned $\sigma_{\mbox{\tiny{NT}}}$ would be larger on average. A gas temperature $T_K = 10$\,K rather than 15\,K  would increase the average \dia\, $\sigma_{\mbox{\tiny{NT}}}\,/\,c_s$ ratio in Oph B by $\sim 20 - 25$\,\%. This increase, while significant, is not large enough for the mean \dia\, $\sigma_{\mbox{\tiny{NT}}}\,/\,c_s$ to match the \amm\, results in Oph B.  With both a constant temperature or decreasing temperature with density, non-thermal motions in Oph B1 and B3 would remain subsonic, while Oph B2 would still be dominated by transonic non-thermal motions. Alternatively, it is possible that the two Class I protostars in Oph B2 could raise the temperature of the dense gas above that traced by \amm\, (no $T_K$ difference was found between protostars and starless areas in \one). In this case, the returned \dia\, $\sigma_{\mbox{\tiny{NT}}}$ would be smaller, and would further increase the differences in magnitude of non-thermal motions seen between the \amm\, and \dia\, emission in Oph B1. In B2, a temperature of 20\,K would decrease the mean $\sigma_{\mbox{\tiny{NT}}}\,/\,c_s$ ratio to 1.1. In a study of dust temperatures in the Ophiuchus Cores, \citet{stamatellos07} showed, however, that embedded protostars in the Cores will only heat very nearby gas and will not raise the mean temperature of the Cores by more than $\sim 1-2$\,K, so an average $T_K = 20$\,K over all Oph B2 is unlikely.  

\subsubsection{Double peaked line profiles}
\label{sec:dblpeak}

\begin{figure}
\plotone{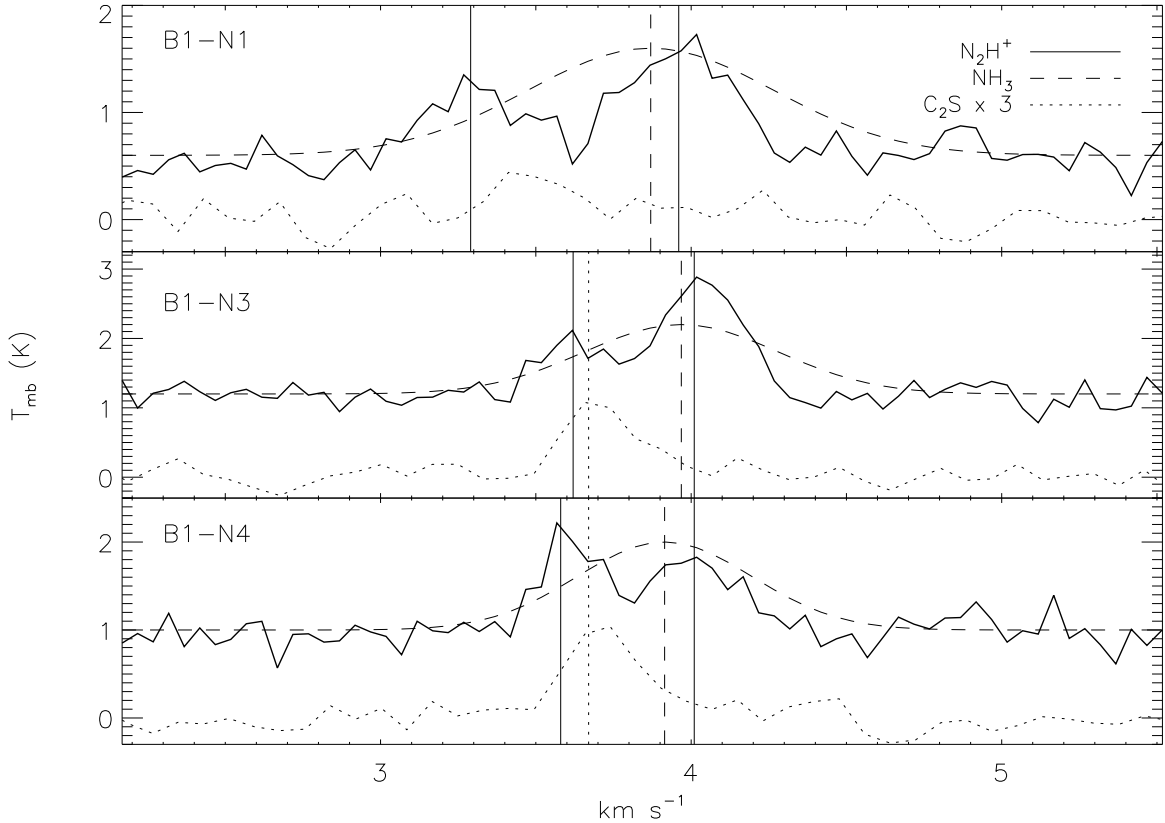}
\caption{Solid black lines show the isolated $F_1F \rightarrow F'_1F' = 0 1 \rightarrow 1 2$ component of the \dia\, 1-0 emission line towards the three \dia\, clumps in Oph B1 which show a double peaked line profile. The dashed curves show the Gaussian fit to \amm\, (1,1) emission, observed wih the GBT, at the \dia\, clump peak location (observations described in \one). The dotted lines show the C$_2$S 2-1 emission at the \dia\, clump peak, also observed with the GBT. Note no significant C$_2$S emission was seen towards B1-N1. The \amm\, Gaussian fit peaks towards the red \dia\, component, while the C$_2$S emission peaks towards the blue \dia\, component. Vertical lines matching each species spectrum show the locations of the fitted $v_{\mbox{\tiny{LSR}}}$ for all species, including both velocity components from the two-component \dia\, 1-0 fit.}
\label{fig:compare_peaks}
\end{figure}

In Figure \ref{fig:compare_peaks}, we show the isolated $F_1F \rightarrow F'_1F' = 0 1 \rightarrow 1 2$ component of the \dia\, 1-0 emission line towards the three \dia\, clumps in Oph B1 (B1-N1, B1-N3, and B1-N4) which show double peaked line profiles, as described in \S\ref{sec:analysis}. Figure \ref{fig:compare_peaks} also shows a single Gaussian line profile overlaid on each \dia\, spectrum, which represents the $v_{\mbox{\tiny{LSR}}}$ and $\Delta v$ of the 18 component Gaussian fit to the \amm\, (1,1) emission at the clump peak, normalized to a peak amplitude of 1\,K. We plot the Gaussian fit for clarity since there are no isolated components in the \amm\, (1,1) hyperfine structure. Additionally, spectra of C$_2$S $2_1 - 1_0$ emission line at the peak clump locations are shown. Note that over all Oph B, C$_2$S emission was only detected towards the southern tip of Oph B1, and thus no significant C$_2$S emission was found at the peak location of B1-N1. The C$_2$S emission was observed with the GBT at $\sim 32$\arcsec\, spatial resolution and 0.08\,\kms\, spectral resolution, with the observations and analysis described in \one. Arrows show the locations of the fitted $v_{\mbox{\tiny{LSR}}}$ for all species, including both \dia\, velocity components. 

The clumps which show the double-peaked line profile structure are the clumps with the highest total \dia\, 1-0 line opacities based on a single velocity component fit, with $\tau = 5, 6$ and 9, respectively for B1-N1, B1-N3 and B1-N4. The $F_1F \rightarrow F'_1F' = 0 1 \rightarrow 1 2$ component is expected to have an opacity $\tau_{iso} = 1/9 \,\tau$, so $\tau_{iso} = 0.6, 0.7$, and $\sim 1$ for the three clumps, based on a single component fit. If the line is optically thick, however, this fit is likely to underestimate the true opacity given the missing, self-absorbed flux. This is suggestive that the lines are indeed self-absorbed. Since the \amm\, and C$_2$S emission is optically thin (\one), however, if the \dia\, emission was self-absorbed due to high optical depth we would expect the emission peak of both \amm\, and C$_2$S to be found between the two \dia\, components in $v_{\mbox{\tiny{LSR}}}$, which is not the case. In all three clumps, we find the $v_{\mbox{\tiny{LSR}}}$ of the \amm\, emission more closely matches the $v_{\mbox{\tiny{LSR}}}$ of the red \dia\, component, while the C$_2$S emission in two clumps ($v_{\mbox{\tiny{LSR}}} =  3.67$\,\kms) more closely matches the $v_{\mbox{\tiny{LSR}}}$ of the blue \dia\, component ($v_{\mbox{\tiny{LSR}}} = 3.85$\,\kms\, and 3.62\,\kms\, for B1-N3 and B1-N4, respectively). This behaviour suggests the \dia\, 1-0 emission is not self-absorbed towards these positions. A similar offset of C$_2$S emission relative to \dia\, was found by \citet{swift06} towards the starless Core L1551, however the case of L1551 the C$_2$S emission is redshifted with respect to the systemic velocity of the Core rather than blueshifted, as we find in B1.

The GBT \amm\, (1,1) observations described in \one\, found extensive emission around Oph B, such that it is possible to determine the $v_{\mbox{\tiny{LSR}}}$ and $\Delta v$ of \amm\, beyond where continuum emission traced the Oph B Core. We use an intensity threshold of 2\,K in the \amm\, (1,1) main component to separate Core and off-Core gas. The off-Core gas surrounding Oph B1 has a mean $v_{\mbox{\tiny{LSR}}} = 3.72$\,\kms\, with an rms variation of 0.13\,\kms, which is significantly different from the mean B1 Core $v_{\mbox{\tiny{LSR}}} = 3.96$\,\kms\, and rms variation of 0.15\,\kms. Both the C$_2$S emission and the blue \dia\, component are thus more kinematically similar to the off-Core \amm\, gas. The critical densities of C$_2$S $2_1 - 1_0$ and \amm\, (1,1) are similar \citep{suzuki92,rosolowsky08}. Kinematically, we find $\sigma_{\mbox{\tiny{NT}}}\,/\,c_s = 0.7$ for the C$_2$S emission if we assume the $T_K$ determined from \amm\, emission accurately represents the temperature of the gas traced by C$_2$S. This result closely matches the \dia\, results in Oph B1, but the non-thermal motions in the gas traced by \amm\, are significantly larger (\S\ref{sec:trends} and \one). It is impossible to match the non-thermal C$_2$S and \amm\, motions for a given $T_K$, since the \amm\, (1,1) line $\sigma_{\mbox{\tiny{NT}}} > \sigma_{obs}$ of the C$_2$S. 

At densities $n \gtrsim$ a few $\times 10^3 - 10^4$\,\cc, chemical models predict C$_2$S is quickly depleted from the gas phase \citep[timescale $t_{dep} \sim 10^5$\,yr;][]{millar90,bergin97}, and accordingly the molecule is observed to be a sensitive tracer of depletion in isolated cores \citep{lai00,tafalla06}. The volume density in Oph B1 is greater than that required for C$_2$S depletion ($n \gtrsim 10^5 - 10^6$\,\cc\, from \S\ref{sec:nex} and continuum observations), so we would not expect to see any C$_2$S emission unless the gas has only been at high density for $t < t_{dep}$. C$_2$S was only detected towards southern B1. An explanation, therefore, for both the presence of C$_2$S in southern B1 and for its velocity offset relative to the dense gas is that gas from the ambient molecular cloud is accreting onto B1, reaching a density high enough to excite the C$_2$S emission line. This material would be chemically `younger' than the rest of the B1 Core gas, such that the C$_2$S has not had enough time to deplete from the gas phase. A similar result was found in a multi-species study of the starless L1498 and L1517B cores, where \citet{tafalla04} conclude that non-spherical contraction of the cores produced asymmetric distributions of CS and CO, where the CS and CO `hot spots' reveal the distribution of recently accreted, less chemically processed dense gas.

A single 850\,\micron\, continuum clump \citep[162715-24303;][]{jorgensen08} was identified in southern Oph B1 where we find B1-N3 and B1-N4. The authors found a clump mass $M = 0.2\,M_\odot$, which is a factor $\sim 2$ less than the virial mass $M_{vir} = 0.4\,M_\odot$ we calculate based on the mean line width (of the single velocity component fit) of B1-N3 and B1-N4 ($\Delta v = 0.52$\,\kms, or $\sigma = 0.22$\,\kms) and the continuum clump radius. This suggests that the clump is currently stable against collapse. If the clump is gaining mass through ongoing accretion from the ambient gas, however, then eventually the non-thermal support may not be large enough to prevent gravitational collapse, leading to the formation of a low-mass protostar. 


\subsubsection{Opacity and Excitation Temperature}

We show in Figure \ref{fig:compare_nh3}c and \ref{fig:compare_nh3}d the distribution of total line opacity $\tau$ and excitation temperature $T_{ex}$ derived from hyperfine line fitting of \amm\, (1,1) and \dia\, 1-0 in Oph B. We find higher total opacities in \dia\, emission ($\langle \tau \rangle = 2.5$) than in \amm\, emission ($\langle \tau \rangle = 1.0$) in Oph B. The \dia\, emission is characterized by lower excitation temperatures than those found for \amm\, emission in \one, with a mean $T_{ex} = 7.1$\,K for \dia\, 1-0 compared with $T_{ex} = 9.5$\,K for \amm\, (1,1). 

\subsubsection{Fractional Abundances and Chemical Evolution}
\label{sec:frac}

We next compare the column densities of \amm\, and \dia\, towards the Oph B Core. We only show results for pixels which have well-determined values for both $N(\mbox{\amm})$ and $N(\mbox{\dia})$. In \one, we found a trend of decreasing fractional \amm\, abundance with increasing $N(\mbox{H$_2$})$ for {\it both} B1 and B2.  After convolving the \amm\, data to match the single-dish \dia\, resolution, we find the following linear relationships between $\log X(\mbox{\amm})$ and $\log N(\mbox{H$_2$})$:

\begin{eqnarray}
\mbox{B1}&:& \log X(\mbox{\amm}) = (5 \pm 4) - (0.6 \pm 0.2) \log N(\mbox{H$_2$}) \\
\mbox{B2}&:& \log X(\mbox{\amm}) = (3.5 \pm 2.2) - (0.5 \pm 0.1) \log N(\mbox{H$_2$})
\end{eqnarray}

\noindent The slopes found for Oph B1 and B2 are both negative and similar to that found in Equation \ref{eqn:xvsh} for the \dia\, fractional abundance as a function of $N(\mbox{H$_2$})$. This result is suggestive of a trend of increasing depletion of both \amm\, and \dia\, with increasing $N(\mbox{H$_2$})$ in Oph B. 

\begin{figure}
\plotone{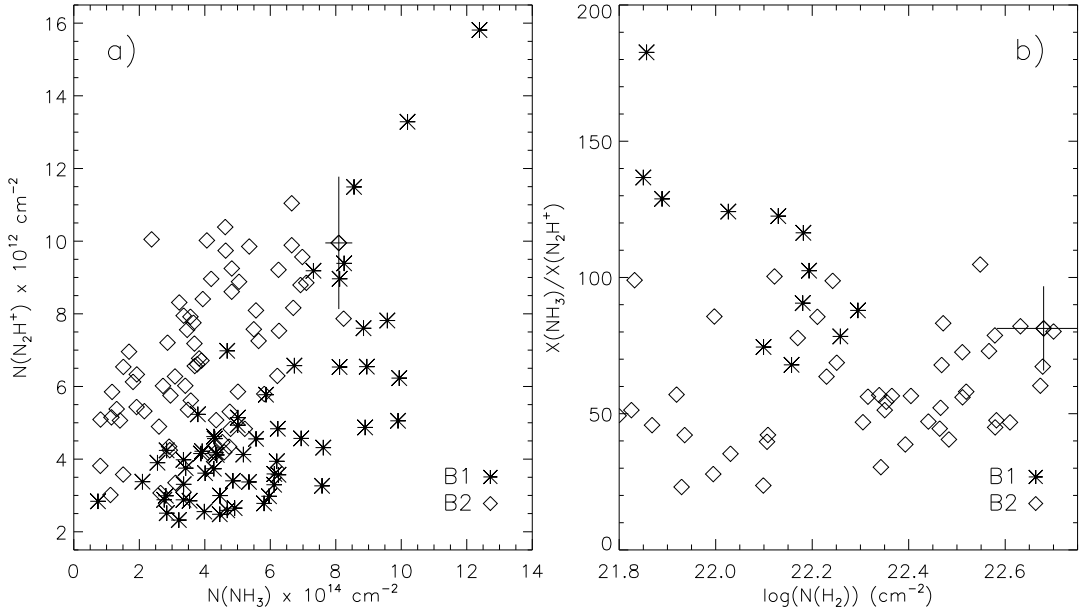}
\caption{a) $N(\mbox{\dia})$ versus $N(\mbox{\amm})$ in Oph B1 and Oph B2. In all plots, the \amm\, data have been convolved to a final 18\arcsec\, FWHM beam, and both the \amm\, and \dia\, data were regridded to 18\arcsec\, pixels. Only pixels where both \amm\, and \dia\, column densities were calculated are plotted. While the mean $N(\mbox{\amm})$ peak at similar values for both B1 and B2, a distribution to higher column values is seen in B1. On average, larger $N(\mbox{\dia})$ values are found in B2. The combination of the spread to higher $N(\mbox{\amm})$ and the lower $N(\mbox{\dia})$ in lead to a larger $N(\mbox{\amm})\,/\,N(\mbox{\dia})$ ratio in B1 than in B2. b) The ratio of \amm\, and \dia\, column densities, $N(\mbox{\amm})\,/\,N(\mbox{\dia})$ plotted as a function of $N(\mbox{H$_2$})$ in Oph B1 and Oph B2. Note that significantly fewer pixels in B1 also have continuum emission above our threshold flux value and are thus not plotted.}
\label{fig:compare_column}
\end{figure}

We show in Figure \ref{fig:compare_column} the \amm\, column density, $N(\mbox{\amm})$, versus the \dia\, column density, $N(\mbox{\dia})$ in Oph B1 and B2, and also the ratio of $N(\mbox{\amm})$ to $N(\mbox{\dia})$ as a function of $N(\mbox{H$_2$})$, calculated in \S\ref{sec:nh2} (we do not show results in B3 due to the small number of pixels with both well-determined \amm\, and \dia\, column densities, but note mean values below). Note that the column density ratio, $N(\mbox{\amm})\,/\,N(\mbox{\dia})$, is equivalent to the ratio of fractional abundances, $X(\mbox{\amm})\,/\,X(\mbox{\dia})$. Figure \ref{fig:compare_column}a shows a significant difference in the $N(\mbox{\amm})\,/\,N(\mbox{\dia})$ ratio between Oph B1 and B2. We find the mean and the standard deviation of this ratio are each twice as large in B1 ($N(\mbox{\amm})\,/\,N(\mbox{\dia}) = 135$, $\sigma = 52$) as in B2 ($N(\mbox{\amm})\,/\,N(\mbox{\dia}) = 65$, $\sigma = 31$) and B3 ($N(\mbox{\amm})\,/\,N(\mbox{\dia}) = 80$, $\sigma = 56$).  In Oph B1 and B2, Figure \ref{fig:compare_column}a illustrates that this difference is largely driven by lower $N(\mbox{\dia})$ values towards B1, while the spread in the $N(\mbox{\amm})\,/\,N(\mbox{\dia})$ appears due to a wider spread of $N(\mbox{\amm})$ values in B1. We find no variation in the $N(\mbox{\amm})\,/\,N(\mbox{\dia})$ ratio as a function of $N(\mbox{H$_2$})$. 

In the clustered star forming region IRAS 20293+3952, \citet{palau07} found strong \amm\, and \dia\, differentiation associated with the presence or lack of embedded YSOs. Low $N(\mbox{\amm})\,/\,N(\mbox{\dia}) \sim 50$ values were found near an embedded YSO cluster, while higher $N(\mbox{\amm})\,/\,N(\mbox{\dia}) \sim 300$ were found towards cores with no associated YSOs. In a study of two dense cores, each containing a starless main body and a YSO offset from the core center, \citet{hotzel04} find a factor $\sim 2$ smaller $X(\mbox{\amm})\,/\,X(\mbox{\dia})$ values towards the starless cores compared with \citeauthor{palau07}, with abundance ratios ($\sim 140-190$, towards the starless gas, and $\sim 60-90$ towards the YSOs) similar to those found in this study.  The relative abundance variation found by \citeauthor{hotzel04} is driven by a varying \amm\, abundance while $X(\mbox{\dia})$ remains constant over the cores. Although we find both a decrease in \dia\, column density as well as an increase in \amm\, column density drives the greater fractional \amm\, abundance towards Oph B1 (see Figure \ref{fig:compare_column}), both $X(\mbox{\dia})$ and $X(\mbox{\amm})$ are greater in B1, by factors of $\sim 2$ and $\sim 4$, respectively, compared with B2 given the lower H$_2$ column densities found in B1 in \S\ref{sec:nh2}. In a survey of 60 low mass cloud cores, \citet{caselli02} find correlations between $N(\mbox{\amm})$ and $N(\mbox{\dia})$ in starless cores, but the relative column density values do not vary significantly between starless and protostellar objects. 

Based on this work and the studies described above, it appears that the relative fractional abundance of \amm\, to \dia\, remains larger towards starless cores than towards protostellar cores by a factor of $\sim 2 - 6$ in both isolated and clustered star forming regions. \edits{We find a 1-$\sigma$ positive trend in the relative \amm\, to \dia\, abundance in Oph B2 with increasing distance to an embedded protostar, suggestive of a direct impact by the protostars on the relative abundances. In their work on x-ray emission from young protostars in Ophiuchus, \citet{casanova95} show that protostellar x-ray emission is potentially significant enough to increase the ionization fraction in nearby dense gas, potentially affecting the local gas chemistry. In this region, the complicated line structures found in both species near the protostars make accurate column density measurements difficult, however, with resulting large uncertainties. }

In their models of collapsing prestellar cores, \citet{aikawa05} found that \amm\, can be enhanced relative to \dia\, in core centers at high densities ($n = 3 \times 10^5 - 3 \times 10^6$\,\cc) due to dissociative recombination reactions of \dia\, and $e^-$ to form NH and N. NH then reacts with H$_3^+$ and H$_2$ to form \amm. The \dia\, recombination reaction is dominant only where CO is depleted (if CO is abundant, \dia\, is destroyed mainly through proton transfer to CO), and occurs when the abundance ratio of CO to electrons, $n(\mbox{CO})\,/\,n_e \lesssim 10^3$. When a collapsing core reaches higher central densities ($n \sim 10^7$\,\cc), however, \citeauthor{aikawa05} predict that the $N(\mbox{\amm})$ will begin to decrease towards the core centre due to depletion, while $N(\mbox{\dia})$ continues to increase, leading to a higher fractional \dia\, abundance relative to \amm\, at later times. This prediction is in agreement with $X(\mbox{\amm})/X(\mbox{\dia})$ results since B2 is more evolved than B1, having already formed protostars, but seems inconsistent with the common slope we find for the decrease in $X(\mbox{\amm})$ and $X(\mbox{\dia})$ with $N(\mbox{H$_2$})$ in both Cores. Detailed modelling of the physical and chemical structure of the (non-spherical and clumpy) Cores is beyond the scope of this study, but would help to constrain the central density and abundance structure as a function of Core radius needed for a direct comparison with the \citeauthor{aikawa05} models. 


\subsection{Are \dia\, 1-0 and \amm\, (1,1) tracing the Oph B Core interior?}

The hyperfine structure of the \amm\, (1,1) inversion transition allows the calculation of volume density $n_{ex}$ as in Equation \ref{eqn:nex}, with the opacity of a typical hyperfine component $\tau_{\mbox{\tiny{\it hf}}} = 0.233 \tau$. The resulting mean density $n_{ex, \mbox{\tiny{\amm}}} = 7.8 \times 10^3$\,\cc\, and $n_{ex, \mbox{\tiny{\amm}}} = 1.7 \times 10^4$\,\cc\, for Oph B1 and B2, respectively (note that these values are slightly less than reported in \one\, due to using the larger 18\arcsec\, beam FWHM). These volume densities  are a factor $\gtrsim 20$ less than those calculated in \S\ref{sec:nex} from \dia\, 1-0 emission ($\sim 26$ in B1 and $\sim 18$ in B2). There is some question whether the \amm\, volume densities are accurate, as recent studies \citep[\one; ][]{foster09} have found \amm-derived $n_{\mbox{\tiny{ex}}}$ values which are an order of magnitude less ($10^4$\,\cc\, versus $10^5$\,\cc\, and greater) than volume densities determined from continuum emission. One possible source of error is the collisional de-excitation rate coefficient, $\gamma_{ul}$, which \citeauthor{foster09} note is reported in various publications with factor of $\sim 10$ variation. Since $n_{cr} = A_{ul}\,/\,\gamma_{ul}$, the critical density is therefore also suspect within a factor $\sim 10$. 

It is clear that in Oph B \amm\, and \dia\, are tracing different motions in the gas. The gas traced by \dia\, emission is significantly more quiescent, as has been observed in high density gas in other star forming cores. The fact that higher densities are calculated from \dia\, emission relative to those calculated from \amm\, emission is thus reasonable, and bolsters the hypothesis that \amm\, (1,1) emission is not tracing the highest density gas. 

\edits{How well does the \dia\, 1-0 emission trace the highest density gas?} In agreement with results in \one, we find that the clumps found in \dia\, 1-0 emission also do not match well continuum clump locations, and significant offsets between \dia\, 1-0 and continuum emission are also apparent in the integrated intensity maps. In particular, both the single-dish- and interferometer-integrated \dia\, 1-0 intensity towards the continuum clump B2-MM8 peak offset to the clump, and in the interferometer data we see an integrated intensity minimum at the MM8 location. On larger scales, we find a trend of {\it decreasing} \dia\, abundance with increasing H$_2$ column density, described above, with a slope in Oph B2 that agrees within uncertainties with that determined for $N(\mbox{\amm})$ versus $N(\mbox{H$_2$})$ for both Oph B1 and B2 (see \S\ref{sec:trends} and \S\ref{sec:frac}). \edits{Furthermore, the mean $X(\mbox{\dia})$ for continuum clumps alone in Oph B is lower than the $X(\mbox{\dia})$ over the whole core (see \S\ref{sec:compare_clumps}). Together, these findings suggest that \dia\, 1-0, though doing a better job than \amm\, (1,1), may not itself trace well the coldest, densest material in Oph B. Appropriate caution must be made when using even \dia\, 1-0 to probe dense gas in some star-forming regions.}

\section{Summary}

We have presented Nobeyama Radio Telescope (18\arcsec\, FWHM) and Australia Telescope Compact Array ($8\arcsec\, \times 5\arcsec$ FWHM) \dia\, 1-0 observations of the Ophiuchus B Core. Our main results are summarized below. 

1. The integrated \dia\, 1-0 intensity is smooth at 18\arcsec\, resolution, and follows generally the 850\,\micron\, continuum emission in the Core as expected from studies of isolated regions, but significant offsets are found between continuum emission and \dia\, integrated intensity peaks. \dia\, clumps identified through the 3D {\sc clumpfind} algorithm correlate well with those previously found in \amm\, emission, but do not correlate well with continuum objects. Closer correspondence between continuum and \dia\, clump locations is found in Oph B1 compared with B2, suggesting \dia\, is better tracing dense gas in this sub-Core. 

2. We find small-scale structure in B2 through high resolution ATCA \dia\, 1-0 data. Towards the continuum object B2-MM8, we find a broken ring of integrated \dia\, emission suggestive of \dia\, depletion. The line widths of the \dia\, emission at high resolution are universally narrower by factors $\gtrsim 2$ than seen in the single-dish data. Some small scale \dia\, structures show velocity gradients which, if due to rotation, are substantially larger than typically seen in star forming cores at lower spatial resolution, but are similar to other interferometric results. 

3. Over all H$_2$ column densities, line widths in Oph B2 are dominated by transonic non-thermal line widths, while non-thermal motions Oph B1 are subsonic. This result shows that \dia\, and \amm\, are not tracing the same material in Oph B, as \amm\, line widths were shown in \one\, to be large and supersonic in both Cores. Attempting to determine a kinetic temperature by forcing equal non-thermal contributions to the \amm\, and \dia\, line widths produces unphysical results. 

4. We find double-peaked \dia\, line profiles in Oph B1. The blue line component is found to match the $v_{LSR}$ of the ambient cloud as traced by \amm\, off-Core emission, while the $v_{LSR}$ of the red line component matches the \amm\, Core gas. The $v_{\mbox{\tiny{LSR}}}$ of optically thin C$_2$S emission also appears to match better that of the ambient cloud, suggesting that Oph B1 is accreting gas from its surroundings. Accretion may cause the co-located continuum clump (which appears itself to be gravitationally stable) to become unstable and form a low mass protostar. 

5. We find a larger \amm\, fractional abundance relative to \dia\, in starless Oph B1 compared with protostellar Oph B2. This result is in agreement with other studies. Chemical models of collapsing cores produce relatively high $N(\mbox{\amm})$ at moderate densities ($n \sim 10^{5-6}$\,\cc) and relatively high $N(\mbox{\dia})$ at high densities ($n \sim 10^{6-7}$\,\cc), in agreement with our results.

6. We find a trend of decreasing fractional \dia\, abundance with increasing H$_2$ column density, as traced by 850\,\micron\, continuum emission. Together with relatively low $X(\mbox{\dia})$ at the positions of the continuum clumps from the single-dish data and the absence of \dia\, emission at the positions of continuum clumps from the interferometer data, this behaviour calls into question the utility of using \dia\, (and \amm) to trace the coldest and densest locations of star-forming regions.


\acknowledgments

We thank the referee, Derek Ward-Thompson, for thoughtful comments which significantly improved the paper. We thank H. Kirk for providing SCUBA maps of the regions observed. JDF thanks Philippe Andr\'{e}, Philip C. Myers and Zhi-Yun Li for helpful discussions that led to the acquisition of the BIMA data presented here. We thank the observatory staff at the ATCA and Nobeyama Radio Observatory for their assistance in making our observations successful. NRO is a branch of the National Astronomical Observatory, National Institutes of Natural Sciences, Japan. The Australia Telescope is funded by the Commonwealth of Australia for operation as a National Facility managed by CSIRO. The James Clerk Maxwell Telescope is operated by The Joint Astronomy Centre on behalf of the Science and Technology Facilities Council of the United Kingdom, the Netherlands Organisation for Scientific Research, and the National Research Council of Canada. The Berkeley-Illinois-Maryland Association (BIMA) array was operated with support from the U.S. National Science Foundation under grants AST 99-81308 to the University of California, Berkeley, AST 99-81363 to the University of Illinois, and AST 99-81289 to the University of Maryland. RKF acknowledges financial support from the University of Victoria and the National Research Council Canada Graduate Student Scholarship Supplement Program. We also acknowledge the support of the National Science and Engineering Research Council of Canada through its Discovery Grant program. 

{\it Facilities:} \facility{ATCA}, \facility{Nobeyama}, \facility{BIMA}

\begin{deluxetable}{llccl}
\tablecolumns{5}
\tablewidth{0pt}
\tablecaption{ATCA Targets in Oph B2 \label{tab:atca}}
\tablehead{
\colhead{R.A.} & \colhead{decl.} & \colhead{rms} & \colhead{Mosaic} & \colhead{Associated Objects}\\
\colhead{J2000} & \colhead{J2000} & \colhead{Jy\,beam$^{-1}$} & \colhead{} & \colhead{}}
\startdata
16:27:24.6 & -24:26:55 & 0.13 & N & B2-A5\tablenotemark{a} \\
16:27:29.3 & -24:27:25 & 0.04 & Y & B2-A6\tablenotemark{a}, B2-MM8\tablenotemark{b},162729-24274\tablenotemark{c}, Elias 32/VSSG18\tablenotemark{d} \\
16:27:31.3 & -24:27:46 & 0.05 & Y & B2-A9\tablenotemark{a}\\
16:27:32.6 & -24:26:56 & 0.12 & N & B2-A7\tablenotemark{a}, B2-MM15\tablenotemark{b} \\
16:27:34.7 & -24:26:13 & 0.10 & Y & B2-A8, B2-A10\tablenotemark{a}, B2-MM16\tablenotemark{b}, 162733-24262\tablenotemark{c}\\
\enddata
\tablenotetext{a}{\citet{friesen09}}
\tablenotetext{b}{\citet{motte98}}
\tablenotetext{c}{\citet{jorgensen08}}
\tablenotetext{d}{\citet{elias78,vssg}}
\end{deluxetable}

\begin{deluxetable}{lccccccc}
\tablecolumns{8}
\tablewidth{0pt}
\tablecaption{\dia\, 1-0 {\sc clumpfind} peaks and parameters in Oph B \label{tab:clump}}
\tablehead{
\colhead{ID} & 
\colhead{RA} & \colhead{decl.} & 
\colhead{FWHM ($x \times y$)} &
\colhead{$T^*_{A,iso}$} & \colhead{$T^*_{A,main}$} & \colhead{$S/N_{iso}$} \\
\colhead{} & \colhead{J2000} & \colhead{J2000} & 
\colhead{(AU)} & \colhead{(K)} & \colhead{(K)} & \colhead{}}
\startdata
B1-N1  &  16 27 13.1 & -24 29 40.3 & 4500 $\times$ 4300 & 1.45 & 2.58 & 7.3  \\  
B1-N2  &  16 27 14.2 & -24 29 15.3 & 5600 $\times$ 4400 & 1.53 & 2.72 & 9.4  \\  
B1-N3  &  16 27 15.3 & -24 30 20.5 & 4400 $\times$ 4800 & 2.28 & 3.01 & 10.9 \\  
B1-N4  &  16 27 16.4 & -24 30 30.5 & 5900 $\times$ 6600 & 1.53 & 2.63 & 7.2  \\ \hline
B12-N1  &  16 27 13.5 & -24 28 10.1 & 6400 $\times$ 6000 & 1.17 & 2.00 & 7.3  \\  
B12-N2  &  16 27 15.3 & -24 27 50.1 & 6000 $\times$ 6000 & 1.36 & 2.23 & 9.4  \\ 
B12-N3  &  16 27 19.0 & -24 27 20.0 & 5600 $\times$ 4900 & 1.05 & 1.85 & 7.0  \\  \hline
B2-N1   &  16 27 21.6 & -24 27 45.1 & 4800 $\times$ 5300 & 0.99 & 1.86 & 7.3  \\  
B2-N2   &  16 27 24.9 & -24 27 15.0 & 4700 $\times$ 6000 & 1.75 & 2.61 & 12.6  \\  
B2-N3   &  16 27 28.2 & -24 27 40.1 & 5900 $\times$ 3800 & 1.33 & 2.60 & 11.6 \\  
B2-N4   &  16 27 28.6 & -24 26 40.0 & 6500 $\times$ 6200 & 0.95 & 3.07 & 7.1  \\ 
B2-N5   &  16 27 29.7 & -24 27 05.0 & 4800 $\times$ 6000 & 1.84 & 3.46 & 14.5  \\ 
B2-N6   &  16 27 32.3 & -24 27 00.0 & 4500 $\times$ 6900 & 2.42 & 3.43 & 18.5 \\ 
B2-N7   &  16 27 33.0 & -24 27 40.1 & 4600 $\times$ 3500 & 1.22 & 2.45 & 7.6  \\  
B2-N8   &  16 27 33.4 & -24 27 45.1 & 4800 $\times$ 3500 & 1.53 & 2.36 & 9.6  \\  
B2-N9   &  16 27 34.5 & -24 26 04.9 & 7400 $\times$ 5100 & 1.11 & 2.48 & 7.6  \\  \hline
B3-N1 &  16 27 22.0 & -24 24 54.7 & $10 000$ $\times$ 7900 & 1.26 & 2.27 & 8.8  \\  
\enddata
\end{deluxetable}

\begin{deluxetable}{cllllll}
\tablecolumns{6}
\tablewidth{0pt}
\tablecaption{\dia\, 1-0 Line Characteristics in Oph B1, B2 and B3 \label{tab:linefit}}
\tablehead{
\colhead{Core} & \colhead{Value} & \colhead{Mean} & \colhead{rms} & \colhead{Min} & \colhead{Max}}
\startdata
Oph B1	& $v_{\mbox{\tiny{LSR}}}$ (\kms)	& 3.96 & 0.15 & 3.43 & 4.13 \\
		& $\Delta v$ (\kms)	& 0.42 & 0.08 & 0.21 & 0.66 \\
		& $\tau$  			& 3.1   & 2.6   & 0.2    & 13 \\
		& $T_{ex}$  (K) 	& 6.2   & 2.0   & 2.1    & 12.1 \\ \hline
Oph B2 	& $v_{\mbox{\tiny{LSR}}}$ (\kms)	& 4.05 & 0.20 & 3.60 & 4.44 \\
		& $\Delta v$ (\kms)	& 0.68 & 0.21 & 0.30 & 1.30 \\
		& $\tau$  			& 2.0   & 1.1   & 0.2    & 5.3 \\
		& $T_{ex}$  (K) 	& 7.2   & 2.8   & 1.9    & 17.8 \\ \hline
Oph B3 	& $v_{\mbox{\tiny{LSR}}}$ (\kms)	& 3.06 & 0.14 & 2.93 & 3.44 \\
		& $\Delta v$ (\kms)	& 0.27 & 0.04 & 0.20 & 0.35 \\
		& $\tau$  			& 2.3   & 2.2   & 0.2    & 7.6 \\
		& $T_{ex}$  (K) 	& 4.5   & 2.8   & 1.2    & 11.7 \\
\enddata
\end{deluxetable}

\begin{figure}
\plotone{figure10a_10d.ps}
\caption{a) Line velocity $v_{\mbox{\tiny{LSR}}}$ (\kms) across the continuum object B2-MM8. Contours begin at 3.8\,\kms\, and increase by 0.1\,\kms. The Class I protostar Elias 32 is shown. The diameter of MM8 is $\sim 30$\arcsec, shown by the white circle. b) Line FWHM $\Delta v$ (\kms) across MM8. Contours begin at 0.5\,\kms\, and increase by 0.1\,\kms. c) Line velocity $v_{\mbox{\tiny{LSR}}}$ (\kms) across the south-eastern edge of Oph B2, near the Class I protostar (shown). Contours begin at 3.5\,\kms\, and increase by 0.2\,\kms. d) Line FWHM $\Delta v$ (\kms) across the south-eastern edge of B2. Contours begin at 0.4\,\kms\, and increase by 0.2\,\kms.}
\label{fig:b-mm8}
\end{figure}

\clearpage
\begin{deluxetable}{lccccccccc}
\tablecolumns{10}
\tablewidth{0pt}
\tablecaption{Derived Parameters at \dia\, 1-0 Peak Locations \label{tab:peak_dat}}
\tablehead{
\colhead{ID} & 
\colhead{$v_{\mbox{\tiny{LSR}}}$} & 
\colhead{$\Delta v$} & 
\colhead{$\tau$} & 
\colhead{$T_{ex}$} & 
\colhead{$\sigma_{\mbox{\tiny{NT}}}\,/\,c_s$} & 
\colhead{$N(\mbox{\dia})$} & 
\colhead{$N(\mbox{H}_2)$} & 
\colhead{$X(\mbox{\dia})$} & \colhead{$n_{ex}$} \\
\colhead{} & 
\colhead{km\,s$^{-1}$} & 
\colhead{km\,s$^{-1}$} & \colhead{} & 
\colhead{K} & 
\colhead{} & 
\colhead{$10^{12}$\,cm$^{-2}$} & 
\colhead{$10^{22}$\,cm$^{-2}$} & 
\colhead{$10^{-10}$} & \colhead{$10^5$\,\cc}} 
\startdata
  B1-N1&  3.95 (1) &   0.42(2) &  6(2) &    5.1(1) &     0.71(4) &      8(2) &     1.6(3) &     5(1) &     0.8(1) \\
 (1)	&	3.96(1) & 0.35(2) & 0.1(1) & 6.8(1) & 0.60 & \nodata & \nodata & \nodata & \nodata \\
 (2)	& 	3.29(5) & 0.4(1)   & 1(1) &  3.9(1) & 0.65 & \nodata & \nodata & \nodata & \nodata \\
 B1-N2&  3.99(1) &   0.35(1) &  3(1) &    6.4(1) &     0.58(2) &      5(2) & \nodata  & \nodata &     1.7(2) \\
 B1-N3&  3.97(1) &   0.50(2) &  5(1) &    5.7(1) &     0.90(3) &    10(2) &     1.6(3) &     6(1) &     1.2(1) \\
 (1)	& 	4.01(1) & 0.35(1) & 5.4(3) & 6.9(1) & 0.58(1) & 8.4(5) & 1.6(3) & 5(1) & 2.1(1) \\
 (2)	& 	3.58(1) & 0.2(2) & \nodata & \nodata & \nodata & \nodata & \nodata & \nodata & \nodata \\
B1-N4&  3.88(1) &   0.53(3) &  9(2) &    4.5(1) &     1.00(5) &     12(3) &     0.8(2) &    17(3) &     0.6(1) \\
 (1)	& 	4.01(1) & 0.32(1) & 3.5(2) & 6.9(1) & 0.53(1) & 5.0(3) & 0.8(2) & 7(1) & 2.1(1) \\
 (2)	& 	3.62(1) & 0.19(8) & \nodata & \nodata & \nodata & \nodata & \nodata & \nodata & \nodata \\ \hline
B12-N1&  3.92(1) &   0.40(2) &  2(1) &    7.1(1) &     0.67(4) &     3(3) &    \nodata &  \nodata &     2.4(2) \\
B12-N2&  4.10(1) &   0.40(2) &  2(1) &    7.0(1) &     0.67(3) &     4(3) &    \nodata &  \nodata &     2.2(2) \\
B12-N3&  4.03(1) &   0.55(2) &  3(1) &    5.5(1) &     0.96(4) &     5(2) &     0.6(1)  &     9(2)  &     1.2(1) \\ \hline
 B2-N1&  3.83(1) &   0.55(3) &  0.4(6) &   14.1(1) &     1.02(5) &   4(11) &   \nodata &  \nodata &  \nodata \\
 B2-N2&  3.73(1) &   0.54(1) &  4.8(7) &    5.6(1) &     1.00(3) &    10(2) &     4.0(8) &     2.5(5) &     1.2(1) \\
 B2-N3&  4.32(1) &   0.47(1) &  1.4(8) &    9.0(1) &     0.74(*) &      5(3) &     2.1(4) &     2.6(5) &     3.7(*) \\
 B2-N4&  4.11(1) &   0.58(2) &  2.7(6) &    7.0(1) &     1.05(3) &      8(2) &     3.1(6) &     2.6(5) &     2.2(2) \\
 B2-N5&  4.25(1) &   0.50(1) &  3.1(6) &    7.5(1) &     0.89(2) &      9(2) &     3.7(7) &     2.5(5) &     2.7(2) \\
 B2-N6&  4.44(1) &   0.39(1) &  5.2(6) &    6.4(1) &     0.68(1) &      9(1) &     2.7(5) &     3.4(7) &     1.6(1) \\
 B2-N7&  4.10(1) &   0.70(2) &  1.6(7) &    7.7(1) &     1.34(5) &      7(3) &     1.3(3) &     6(1) &     3.4(4) \\
 B2-N8&  4.09(1) &   0.65(2) &  1.6(7) &    7.7(1) &     1.16(4) &      6(3) &     0.8(2) &     9(2) &     2.8(3) \\
 B2-N9&  4.03(1) &   0.61(2) &  2.9(8) &    6.3(1) &     1.13(4) &      8(2) &     2.9(6) &     2.9(6) &     1.7(1) \\ \hline
 B3-N1&  2.99(1) &   0.21(1) &  3(1) &    6.3(1) &     0.20(2) &      2(1) & \nodata &  \nodata &     1(2) \\\enddata
\tablecomments{Uncertainties are given in brackets beside values and show the uncertainty in the last digit of the value. The $N(\mbox{H$_2$})$ and $X(\mbox{\dia})$ uncertainties include the $\sim 20$\% uncertainty in the submillimeter continuum flux values only; uncertainties in $T_d$ and $\kappa_\nu$ are not taken into account.}
\end{deluxetable}

\begin{deluxetable}{cllllll}
\tablecolumns{6}
\tablewidth{0pt}
\tablecaption{Derived Column Densities, Fractional Abundances and Non-thermal Line Widths in Oph B1, B2 and B3 \label{tab:col}}
\tablehead{
\colhead{Core} & \colhead{Value} & \colhead{Mean} & \colhead{rms} & \colhead{Min} & \colhead{Max}}
\startdata
Oph B1	& $N(\mbox{\dia})$ ($\times 10^{12}$\,cm$^{-2}$)	& 4.5 & 2.7 & 2.1 & 15 \\
		& $X(\mbox{\dia})$ ($\times 10^{-10}$)			& 15 & 14 & 3.7 & 54 \\ 
		& $n_{ex}$ ($\times 10^5$\,\cc)					& 2.0 & 2.0 & 0.4 & 8.9 \\
		& $\sigma_{\mbox{\tiny{NT}}}\,/\,c_s$ (\kms)		& 0.71 & 0.16 & 0.38 & 1.16 \\ \hline
Oph B2 	& $N(\mbox{\dia}$) ($\times 10^{12}$\,cm$^{-2}$)	& 5.9 & 1.9 & 1.0 & 9.8 \\
		& $X(\mbox{\dia}$) ($\times 10^{-10}$) 			& 5.6 & 5.4 & 1.6 & 29 \\
		& $n_{ex}$ ($\times 10^5$\,\cc)					& 3.1 & 2.9 & 1.0 & 14 \\
		& $\sigma_{\mbox{\tiny{NT}}}\,/\,c_s$ (\kms)		& 1.26 & 0.40 & 0.64 & 2.3 \\ \hline
Oph B3\tablenotemark{a} 	& $N(\mbox{\dia}$) ($\times 10^{12}$\,cm$^{-2}$)	& 1.9 & 0.4 & 1.2 & 2.9 \\
		& $n_{ex}$ ($\times 10^5$\,\cc)					& 2.0 & 1.5 & 0.2 & 5.1 \\
		& $\sigma_{\mbox{\tiny{NT}}}\,/\,c_s$ (\kms)		& 0.39 & 0.13 & 0.14 & 0.65 \\ 
\enddata
\tablenotetext{a}{The 850\,\micron\, flux at Oph B3 is less than the required threshold of $S_\nu \geq 0.1$\,Jy\,beam$^{-1}$ (\S4.1.2); therefore we do not comment on the fractional \dia\, abundance.}
\end{deluxetable}

\begin{deluxetable}{lccccccccc}
\tablecolumns{10}
\tablewidth{0pt}
\tablecaption{Derived Parameters for 850\,\micron\, clumps and Class I protostars \label{tab:prot_dat}}
\tablehead{
\colhead{ID} & 
\colhead{$v_{\mbox{\tiny{LSR}}}$} & 
\colhead{$\Delta v$} & 
\colhead{$\tau$} & 
\colhead{$T_{ex}$} & 
\colhead{$\sigma_{\mbox{\tiny{NT}}}\,/\,c_s$} & 
\colhead{$N(\mbox{\dia})$} & 
\colhead{$N(\mbox{H}_2)$} & 
\colhead{$X(\mbox{\dia})$} & \colhead{$n_{ex}$} \\
\colhead{} & 
\colhead{km\,s$^{-1}$} & 
\colhead{km\,s$^{-1}$} & \colhead{} & 
\colhead{K} & 
\colhead{} & 
\colhead{$10^{12}$\,cm$^{-2}$} & 
\colhead{$10^{22}$\,cm$^{-2}$} & 
\colhead{$10^{-10}$} & \colhead{$10^5$\,\cc}} 
\startdata
850\,\micron \\ \hline
162712-24290 &  3.66(1) &   1.36(1) &  \nodata &  \nodata &     2.52(1) &  \nodata &  \nodata &  \nodata &  \nodata \\
162713-24295 &  3.86(1) &   0.52(2) &  4(1) &    5.6(1) &     0.90(5) &      8(3) &     2.1(4) &    3.7(7) &     1.1(1) \\
162715-24303 &  3.93(1) &   0.50(2) &  6(1) &    5.5(1) &     0.92(4) &     10(2) &     2.0(4) &    5(1) &     1.1(1) \\
162725-24273 &  3.72(1) &   0.61(2) &  4.7(8) &    5.3(1) &     1.16(4) &   10(2) &     4.1(8) &    2.4(5) &     1.0(1) \\
162728-24271 &  4.18(1) &   0.61(2) &  2.4(6) &    7.9(1) &     1.15(3) &  9(2) &     6(1) &    1.6(3) &     3.4(4) \\
162729-24274 &  4.36(1) &   0.55(1) & \nodata &  \nodata &     0.94(*) &  \nodata &  \nodata &  \nodata &  \nodata \\
162730-24264 &  4.16(1) &   0.57(2) &  1.6(7) &    9.6(1) &     0.94(*) &  8(3) &     2.6(5) &    3.2(6) &     \nodata \\
162733-24262 &  4.23(1) &   0.73(2) &  2.8(7) &    6.3(1) &     1.35(4) &  10(2) &     4.2(8) &    2.3(5) &     1.7(1) \\
\hline
Class I  \\ \hline
Oph-emb 5 &  3.82(1) &   0.58(3) &  1(1) &    8.3(1) &     1.07(5) &      4(5) &  \nodata &   \nodata &     3.9(5) \\
Oph-emb 11 &  \nodata &  \nodata &  \nodata &  \nodata &  \nodata &  \nodata &  \nodata &  \nodata &  \nodata \\
Oph-emb 19 &  4.31(1) &   0.52(1) &  1.9(7) &    8.4(1) &     0.87(*) &      7(3) &     2.3(5) &     3.1(6) &     3.3(*) \\
Oph-em 26 &  3.87(1) &   1.56(1) &  \nodata & \nodata &     2.96(1) &  \nodata &     2.3(5) &   \nodata &  \nodata \\
\enddata
\tablecomments{850\,\micron\, continuum clump locations taken from \citet{jorgensen08}. Class I protostar locations taken from \citet{enoch09}.  Uncertainties are given in brackets beside values and show the uncertainty in the last digit of the value. The $N(\mbox{H$_2$})$ and $X(\mbox{\dia})$ uncertainties include the $\sim 20$\% uncertainty in the submillimeter continuum flux values only; uncertainties in $T_d$ and $\kappa_\nu$ are not taken into account. A (*) indicates that the uncertainty in the value at this position were large, even though in most cases the values themselves are reasonable. }
\end{deluxetable}

\begin{deluxetable}{lccccccccc}
\tablecolumns{10}
\tablewidth{0pt}
\tablecaption{Mean Derived Physical Parameters for \dia\, clumps, 850\,\micron\, clumps and Class I protostars \label{tab:mean_dat}}
\tablehead{
\colhead{} & 
\colhead{$v_{\mbox{\tiny{LSR}}}$} & 
\colhead{$\Delta v$} & 
\colhead{$\tau$} & 
\colhead{$T_{ex}$} & 
\colhead{$\sigma_{\mbox{\tiny{NT}}}\,/\,c_s$} & 
\colhead{$N(\mbox{\dia})$} & 
\colhead{$N(\mbox{H}_2)$} & 
\colhead{$X(\mbox{\dia})$} & \colhead{$n_{ex}$} \\
\colhead{} & 
\colhead{km\,s$^{-1}$} & 
\colhead{km\,s$^{-1}$} & \colhead{} & 
\colhead{K} & 
\colhead{} & 
\colhead{$10^{12}$\,cm$^{-2}$} & 
\colhead{$10^{22}$\,cm$^{-2}$} & 
\colhead{$10^{-10}$} & \colhead{$10^5$\,\cc}} 
\startdata
\dia\, clumps 			& 3.98 & 0.49 & 3.2 & 7.0 & 0.86 & 7.0 & 1.9 & 8.0 & 1.9 \\
850\,\micron\, clumps 	& 3.99 & 0.70 & 3.9 & 6.7 & 1.27 & 9.4 & 3.2 & 3.1 & 1.7 \\
Class I protostars		& 4.00 & 0.89 & 1.5* & 6.5* & 1.67 & 5.8* & 2.3* & 3.1* & 3.6* \\
\enddata
\tablecomments{Only three protostars (two where values labelled with *) were associated with significant \dia\, 1-0 emission such that physical parameters could be determined from the HFS line fitting routine. }
\end{deluxetable}

\bibliographystyle{apj}
\bibliography{biblio}

\end{document}